\documentclass[reprint,superscriptaddress,amsmath,amssymb,aps,prb]{revtex4-2}
\usepackage{graphicx}
\usepackage{dcolumn}
\usepackage{bm}
\usepackage[colorlinks,allcolors=blue]{hyperref}
\usepackage{xcolor}
\usepackage{gensymb}
\usepackage{natbib}
\usepackage{txfonts}
\usepackage[final]{changes}
\usepackage{changes}

\usepackage{orcidlink}

\begin{document}

\preprint{}

\title{Two-dimensional non-van der Waals niobium nitride nanosheets with high-temperature two-gap superconductivity}

\author{Si-Yi Xiong}
\affiliation{Laboratory for Quantum Design of Functional Materials, and School of Physics and Electronic Engineering, Jiangsu Normal
University, Xuzhou 221116, China}

\author{Peng Jiang \orcidlink{0000-0002-4291-608X}}
\email{pjiang@jsnu.edu.cn; pjiang93@mail.ustc.edu.cn}
\affiliation{Laboratory for Quantum Design of Functional Materials, and School of Physics and Electronic Engineering, Jiangsu Normal
University, Xuzhou 221116, China}

\author{Yiming Wang}
\affiliation{Center for High Pressure Science and Technology Advanced Research, Shanghai 201203, China}

\author{Yan-Ling Li \orcidlink{0000-0002-0144-5664}}
\email{ylli@jsnu.edu.cn}
\affiliation{Laboratory for Quantum Design of Functional Materials, and School of Physics and Electronic Engineering, Jiangsu Normal
University, Xuzhou 221116, China}

%\date{\today}

\begin{abstract}

The exploration of the superconductivity in two-dimensional materials has garnered significant attention due to their promising low-power applications and 
fundamental scientific interest.
Here, we report some novel stable non-van der Waals 
Nb$_x$N$_{x+1}$ ($x$ = 1-4) monolayers 
derived from the NbN bulk exfoliated along the [001] direction, 
as identified through first-principles calculations. 
Among these monolayers, Nb$_2$N$_3$, which crystallizes in 
the $P\overline{6}m2$ symmetry, stands out with an exceptional 
superconducting transition temperature of 77.8~K, 
setting a new high-$T_c$ benchmark for two-dimensional transition 
metal nitrides and binary compounds. 
Our detailed analysis reveals that the strong superconductivity in Nb$_2$N$_3$ is 
driven by phonon modes dominated by N vibrations, with significant 
electron-phonon coupling contributions from N-$p$ and Nb-$d$ electronic states.
Using the anisotropic Migdal-Eliashberg framework, we further determine the 
two-gap nature of the superconductivity in the Nb$_2$N$_3$ monolayer, 
characterized by pronounced electron-phonon coupling and anisotropic 
energy gaps. 
These results advance our understanding of superconductivity in two-dimensional transition 
metal nitride and highlight their potential for nanoscale superconducting applications.

\end{abstract}

\maketitle

\section{Introduction}
%%%
Two-dimensional (2D) materials serve as an ideal platform in condensed matter physics for exploring novel fundamental phenomena, including superconductivity \cite{doi:10.1126/science.aab2277,xi2016ising,hadma.202006124}, 
quantum anomalous Hall effect \cite{lu2024fractional,PhysRevLett.132.106602}, 
ferrovalley \cite{tong2016concepts,PhysRevB.104.035430}, photogalvanic effect \cite{2021Two,PhysRevB.102.081402}, and 
topological spin texture \cite{PhysRevB.101.060404} etc.
In particular, 2D superconductors have attracted significant attention due to their 
intriguing and fruitful physical properties, 
such as, topological superconductivity \cite{hsu2017topological,PhysRevB.103.104503}, high upper critical magnetic field \cite{doi:10.1126/science.aab2277},
quantum metallic state \cite{doi:10.1126/science.aax5798}, 
and quantum Griffith singularity \cite{PhysRevLett.127.137001}. 
Compared to three-dimensional counterparts, 2D superconductors generally 
exhibit enhanced tunability \cite{saito2016highly}, positioning them as a 
promising electronic medium for exploring a more rich superconducting properties 
by various external stimuli, such as, strain \cite{PhysRevB.96.094510,PhysRevB.93.155430}, atomic intercalation \cite{PhysRevB.93.155430,PhysRevB.96.235426}, 
interface engineering \cite{peng2014tuning}, and carrier doping \cite{doi:10.1021/acs.nanolett.2c02947}.
Numerous studies on 2D superconductivity predominantly concentrated on van der Waals (vdW) 
layered materials, particularly transition metal dichalcogenides (TMDs), 
due to their novel superconducting properties, along with the favorable exfoliation and structural stability \cite{doi:10.1021/acs.nanolett.9b04891,PhysRevB.99.220503,PhysRevB.104.174510,navarro2016enhanced}. It is studied 
that superconductivity in certain atomically thin TMDs 
has been observed to persist in in-plane magnetic fields significantly excessing 
the Pauli paramagnetic limit \cite{doi:10.1126/science.aab2277,xi2016ising}. This behavior is attributed to Ising pairing where Dresselhaus-type spin–orbit coupling (SOC) constrains the electron spins to align in the out-of-plane direction, 
and mitigating the pair-breaking impact of the in-plane field \cite{hsu2017topological,de2018tuning}. 
Besides, the coexistence of superconductivity and charge density wave phases can
be observed in 2D NbSe$_2$ system \cite{doi:10.1021/acs.nanolett.8b00237}.

Beyond TMDs, numerous studies have identified 
2D transition-metal nitrides (TMNs) as a compelling family of 
materials owing to their rich and highly tunable 
superconducting properties \cite{PhysRevB.105.165101,campi2021prediction,PhysRevB.103.064510}.
A notable example is the exfoliated 2D ZrN, which exhibit a superconducting transition temperature ($T_c$) of 2 K and show a dimensional crossover in 
2D superconductivity, marked by an upper critical field beyond the 
Pauli paramagnetic limit \cite{guo2019freestanding}. This behavior
offers valuable insights into the nature 
of non-centrosymmetric superconductivity in the presence of spin and valley 
coupling. On the other hand, theoretical investigations indicate that 
the superconductivity of 2D TMNs can be significantly improved by some external 
control means. For example, the $T_c$ of the monolayer Ba$_2$N is 
predicted to be 3.4~K and could increase to 10.8~K
via applying a tensile strain due to the enhanced inner-layer electron-phonon coupling (EPC)
and noticeable phonon softening \cite{PhysRevB.105.165101}.
It has also been demonstrated that applying 2\% biaxial strain to 
a superconducting Mo$_2$N monolayer triggers a new superconducting gap, 
which is further enhanced with the increasing strain, thereby opening 
additional coupling model between strain and superconducting gap \cite{D2NR00395C}.
Furthermore, by passivating N-terminated surfaces of the TaN$_2$ monolayer 
with Si atoms, the $T_c$ can be significantly increased to approximately 24.6 K \cite{yan2021surface}. 
However, achieving a effective strain or other modulation methods in 
experimental settings poses significant challenges, preventing the attainment of the ideal $T_c$. Consequently, exploring more 2D TMNs with intrinsic metallicity 
that exhibit stable and elevated $T_c$ for practical applications 
remains a considerable challenge.

In addition to the aforementioned challenges, synthesizing 2D TMNs presents significant difficulties, leading to most of these materials being obtained through exfoliation 
from vdW layered structures, which exhibit very weak interlayer interactions. 
For instance, the recently reported 2D topological superconductor W$_2$N$_3$ 
is derived from its vdW layered  bulk counterpart \cite{PhysRevB.103.064510,campi2021prediction,PhysRevB.103.104503}.
Furthermore, the $T_c$ of 2D W$_2$N$_3$ is nearly 39~K, 
comparable to that of the well-known conventional 
bulk superconductor MgB$_2$ \cite{doi:10.1073/pnas.67.1.313}, 
indicating that a high $T_c$ may be achievable 
when the ratio of transition metal to nitrogen is 2:3.
Nevertheless, the availability of vdW layered materials is relatively 
limited compared to traditional non-vdW materials, 
which somewhat hinders the exploration of additional 
promising 2D superconductors. 
Recent experimental advances in 2D non-vdW materials, particularly highlighted by 
the demonstration of a new 2D material `hematene' derived from natural 
iron ore hematite ($\alpha$-Fe$_2$O$_3$) via liquid exfoliation, have sparked significant 
interest in monolayer non-vdW materials \cite{puthirath2018exfoliation,jiang2023mechanical}.
Unlike vdW materials, non-vdW monolayers require advanced exfoliation techniques 
due to the stronger chemical bonds that must be 
disrupted to freestanding single layers. This makes them fundamentally different 
from traditional vdW monolayer systems and 
opens new avenues for discovering exotic quantum phases 
in strongly bonded 2D materials \cite{doi:10.1021/acs.nanolett.3c00113,doi:10.1021/acs.nanolett.1c03841,doi:10.1021/acsnano.4c12155}. As a result, the exploration of non-vdW monolayer
superconductors plays a critical role in advancing the search for 
new high-temperature superconductors. 
Moreover, it is emphasized that the successful isolation of 
monolayer non-vdW transition metal oxides suggests that 2D non-vdW TMNs could also be 
obtained through similar synthetic means \cite{puthirath2018exfoliation,jiang2023mechanical}, 
as the weaker bonding of nitrogen relative to oxygen allows for easier exfoliation.

To this end, we here mainly focus on niobium nitride system 
because Nb has a relatively light mass and high $T_c$ at ambient pressure 
among various transition-metal elements. 
By cleaving the non-vdW bulk niobium nitride,
we predict a series of 2D Nb$_x$N$_{x+1}$ ($x$=1-4) nanosheets. 
First-principles calculations demonstrate that the predicted nanosheets 
are dynamically stable verified by the phonon spectra.
Anisotropic Migdal-Eliashberg theory unveils that the monolayer Nb$_2$N$_3$ 
is a two-gap superconductor with large EPC 
and has the highest $T_c$ of about 77.8~K among these nanosheets. 
The $T_c$ value is a record high $T_c$ for 2D TMNs and even binary compounds 
without the help of gating, chemical decoration and external strain 
at ambient condition.

\section{Computational Methods}
The structural and electronic properties of the Nb$_2$N$_3$ monolayer are calculated 
based on the density functional theory (DFT) with projector
augmented wave (PAW) method, as implemented in the Vienna {\it Ab-initio} Simulation Package (VASP) \cite{kresse1996efficiency,kresse1996efficient,kresse1999ultrasoft}. The generalized gradient approximation (GGA) of the Perdew-Burke-Ernzerhof (PBE) form \cite{PhysRevB.88.085117} is 
adopted for the exchange-correlation functional. 
The cut-off energy of the plane wave basis is set to 500~eV. 
All atomic structures are relaxed until the Hellmann-Feynman force on each atom is less than 
1~meV/{\AA}.
To avoid the interactions between the periodic imagines for the monolayer
system, the vacuum layer thickness greater than 15~{\AA} is applied along the out-of-plane direction. 
The 2D Brillouin zone (BZ) is sampled with a {\it $\textbf{k}$} grid of 12$\times$12$\times$1 
by using a Monkhorst-Pack scheme.

The phonon spectra and EPC properties of the 
Nb$_x$N$_{x+1}$ ($x$ = 1-4) monolayers are calculated based on
density functional perturbation theory (DFPT), 
as implemented in the QUANTUM ESPRESSO (QE) package \cite{giannozzi2017advanced,giannozzi2020quantum}, in which
optimized norm-conserving Vanderbilt (ONCV) pseudopotentials are employed
\cite{PhysRevB.88.085117}.
The {\it $\textbf{k}$}-point and {\it $\textbf{q}$}-point grids are set to 16$\times$16$\times$1 and 8$\times$8$\times$1, respectively.
The cut-off energies of the wave function and charge density 
are 90~Ry and 450~Ry, respectively. 
Furthermore, we employed an EPC matrix 
interpolation
with the denser {\it $\textbf{k}$} and {\it $\textbf{q}$} meshes 
of 288$\times$288$\times$1 and 144$\times$144$\times$1 to 
calculate superconducting properties by solving the anisotropic Migdal-Eliashberg equation,
as implemented in the EPW code \cite{PhysRevB.76.165108,PhysRevB.87.024505,ponce2016epw}. 
More computational details of EPC constant and anisotropic Migdal-Eliashberg equations can be found in the Supplemental Material (SM) ~\cite{supp}.
Note that convergence tests were carried out for the plane-wave or charge-density cutoff energies in both VASP and QE, and the selected cutoff values 
are sufficient to ensure the accuracy of the results.

\begin{figure}
        \centering
        \includegraphics[width=1.0\linewidth]{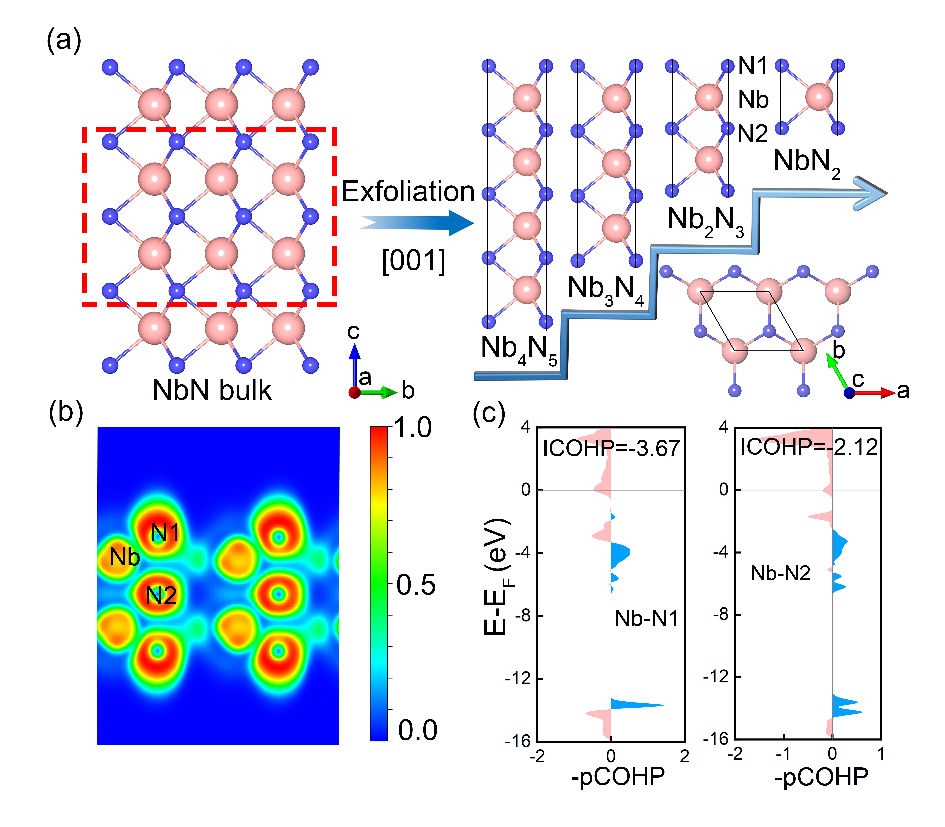}
 \caption{
 (a) Side view of the crystal structure of the bulk NbN (left), the exfoliated nanosheets 
 along the [001] direction is indicated by the dashed squares. Structure of NbN$_2$, Nb$_2$N$_3$, Nb$_3$N$_4$and Nb$_4$N$_5$ monolayer in side views (right). The side views of the atomic structure of the Nb$_2$N$_3$ monolayer (bottom right). The unit cell is indicated by the solid black lines, Nb and N atoms are displayed in pink and blue colors, respectively. 
 (b) 2D slices projected ELF. (c) The pCOHP states of the Nb$_2$N$_3$ monolayer.
    }\label{fig1}
\end{figure}

\section{Results and Discussions}
%\subsection{Structural properties}

Here, we focus exclusively on Nb$_x$N$_{x+1}$ ($x$ = 1-4) nanosheets with N atoms 
serving as terminations, primarily due to the enhanced stability of N atoms 
compared to Nb at the interface. These 2D Nb$_x$N$_{x+1}$ nanosheets, 
featuring a varying number of atomic layers ($n$ = 3-9), can be cleaved from 
the experimentally known tungsten-carbide (WC)-type NbN bulk material
along the [001] direction \cite{goldschmid2013interstitial,PhysRevB.99.104508,adma.201102306},
including NbN$_2$, Nb$_2$N$_3$, Nb$_3$N$_4$ and Nb$_4$N$_{5}$, 
as shown in Fig.~\ref{fig1}(a). 
In previous reports,
some techniques such as liquid-phase exfoliation, molecular beam epitaxy, 
and chemical vapor deposition have been widely applied to the synthesis of 
2D non-vdW materials, such as, $\alpha$-Fe$_2$O$_3$ \cite{puthirath2018exfoliation},
 SnO \cite{jiang2023mechanical}, KV$_3$Sb$_5$ \cite{jiang2023mechanical}, and W$_5$N$_6$ films \cite{doi:10.1021/acsnano.4c12155}.
These suggests that similar approaches could be 
effectively employed for the synthesis of the Nb$_x$N$_{x+1}$ monolayers.
The structural information for all of Nb$_x$N$_{x+1}$ nanosheets are summarized in Table S1 of the SM~\cite{supp}.
First, we calculate the phonon spectra of these nanosheets, as shown in Fig.~S1 \cite{supp}, 
revealing the absence of imaginary frequencies and thus confirming their 
excellent dynamical stability. It is emphasized that we mainly focus on the 
superconducting properties of these novel materials,
and we find that $T_c$ values of the Nb$_2$N$_3$, Nb$_2$N$_3$, Nb$_3$N$_4$ and Nb$_4$N$_{5}$ 
monolayers are all beyond 30~K (see Fig.~S2 \cite{supp}). Interestingly, the Nb$_2$N$_3$ monolayer 
exhibits the highest $T_c$ among them. Therefore, 
we will mainly concentrate on the monolayer Nb$_2$N$_3$ system, 
with a detailed discussion of its superconductivity to follow.

Before exploring the superconducting properties of the Nb$_2$N$_3$ monolayer, 
we first systematically examine its detailed structures [see Fig.~\ref{fig1}(a)].
The Nb$_2$N$_3$ monolayer displays a high-symmetric atomic structure that contains five atomic layers 
with space group $P\overline{6}m$2 (No.187), where each atomic layer contains only one elemental 
species with the atoms in a given layer arranged in a triangular lattice, and then are stacked
in the N-Nb-N-Nb-N sequence with AB stacking type. Each Nb atom is bonded with 
six N atoms that include two inequivalent N atoms (i.e. the outer N1 and inner N2), 
and N2 atoms are
sandwiched by Nb atoms while N1 atoms are located at the surface and bonds to three Nb atoms.
The optimized in-plane lattice constant and the vertical distance between the outer N1 atoms of 
the Nb$_2$N$_3$ monolayer are 2.73~{\AA} and 4.94~{\AA,} respectively.
To demonstrate the potential of the experimental
synthesis of the Nb$_2$N$_3$ monolayer, we first calculate its
formation, that is, 
$E_\mathrm{for}$ = ($E_\mathrm{Nb_2N_3}$-2$E_\mathrm{Nb}$-3$E_\mathrm{N}$)/5,
where $E_\mathrm{Nb}$ and $E_\mathrm{N}$ are the energies of Nb and N in 
their most stable bulk or gas phases, respectively. The calculated $E_\mathrm{for}$ is
-0.57~eV/atom, and its negative value suggests that the fabrication is about
possible. 
We have also considered other three common structural phases ($P\bar{3}1m$, $P\bar{3}m1$ 
and $P6/mmm$) and calculated the corresponding energies, as shown in Fig.~S3 \cite{supp}.
Our results reveal that the predicted $P\bar{6}m2$ phase is not a ground geometry among the considered candidate phases. Specifically, its energy is higher than that 
of the $P\bar{3}m1$ phase but lower than those of 
the $P\bar{3}1m$ and $P6/mmm$ phases, indicating that the predicted $P\bar{6}m2$ 
phase is metastable. It is important to emphasize that metastability does not preclude experimental realization. For instance, in transition metal dichalcogenides 
(e.g., MoS$_2$, WS$_2$ etc.) \cite{https://doi.org/10.1002/adfm.201802473}, both the T ($P\bar{3}m1$) 
and H ($P\bar{6}m2$) phases have been successfully synthesized, 
demonstrating that metastable phases can indeed be experimentally accessible.

The thermal stability of the Nb$_2$N$_3$ monolayer is confirmed by 
AIMD simulations shown in Fig.~S4 \cite{supp}, in which the thermal-induced fluctuation induces
slight changes in the energy and atomic structure.
The mechanical properties of the Nb$_2$N$_3$ monolayer 
are then investigated. In a 2D hexagonal system, the elastic behavior is characterized by two independent elastic constants, $C_{11}$ and $C_{12}$, 
with the relationships $C_{11}$ = $C_{22}$, 
and $C_{66}=(C_{11}-C_{12})/2$ \cite{D2TC04799C}. 
Within the framework of Voigt notation, 
 the strain-dependent energy density for such a system is expressed as 
 $E_\mathbf{s}=\frac{1}{2}C_{11}\varepsilon_{x}^{2}+\frac{1}{2}C_{22}\varepsilon_{y}^{2}+C_{12}\varepsilon_{x}\varepsilon_{y}+2C_{66}\varepsilon_{xy}^{2}$ \cite{PhysRevB.102.195408}, 
where $\varepsilon_{x}$ and $\varepsilon_{x}$ represent the normal strains along $x$ and
$y$ directions, respectively, and $\varepsilon_{xy}$ denotes the shear strain. 
Through the energy-strain method, C$_{11}$, C$_{12}$, and C$_{66}$ are found to be 
365.04~N/m, 169.69~N/m, and 97.68~N/m, respectively.
These values satisfy the stability conditions for a 2D hexagonal lattice, which require
$C_{11} > 0$ and $C_{11} > \left| C_{12}\right|$ \cite{D2TC04799C,haastrup2018computational}.
This confirms the mechanical stability of the Nb$_2$N$_3$ monolayer.

Next, the electron localization function (ELF) and projected crystal orbital Hamilton population (pCOHP) \cite{maintz2016lobster} are calculated to reveal the bonding characteristic of the Nb$_2$N$_3$ monolayer, 
and the results are depicted in Fig.~\ref{fig1}(b) and \ref{fig1}(c), respectively. 
Given that the elemental electronegativity of N exceeds that of Nb, electrons 
predominantly localized around the N atoms. 
The ELF values between N2 and Nb are notably lower than those between N1 and Nb, 
indicating a stronger covalent interaction for the Nb-N1 bond.
The presence of fully occupied bonding states and partially filled 
antibonding states provides compelling evidence for the covalent 
nature of the Nb-N1 and Nb-N2 bonds. Furthermore, the negative integrated COHP (ICOHP) values confirm 
the bonding characteristics between Nb and both N1 and N2. 
Specifically, the ICOHP value for the shorter Nb-N1 bond is -3.67 eV per pair, 
which is more negative than that of the Nb-N2 interactions, 
aligning well with the ELF results.
In addition, we note that the pCOHP exhibits a prominent antibonding character at the Fermi level, and the ICOHP value is negative. This indicates that bonding interactions 
continue to dominate despite the presence of antibonding states near the 
Fermi level. A similar phenomenon has been reported in systems, 
such as, BiH$_4$ \cite{doi:10.1021/jacs.4c15137} and SrHfH$_{18}$ \cite{PhysRevB.109.184516}.

\begin{figure}[thp]
    \includegraphics[width=0.48 \textwidth]{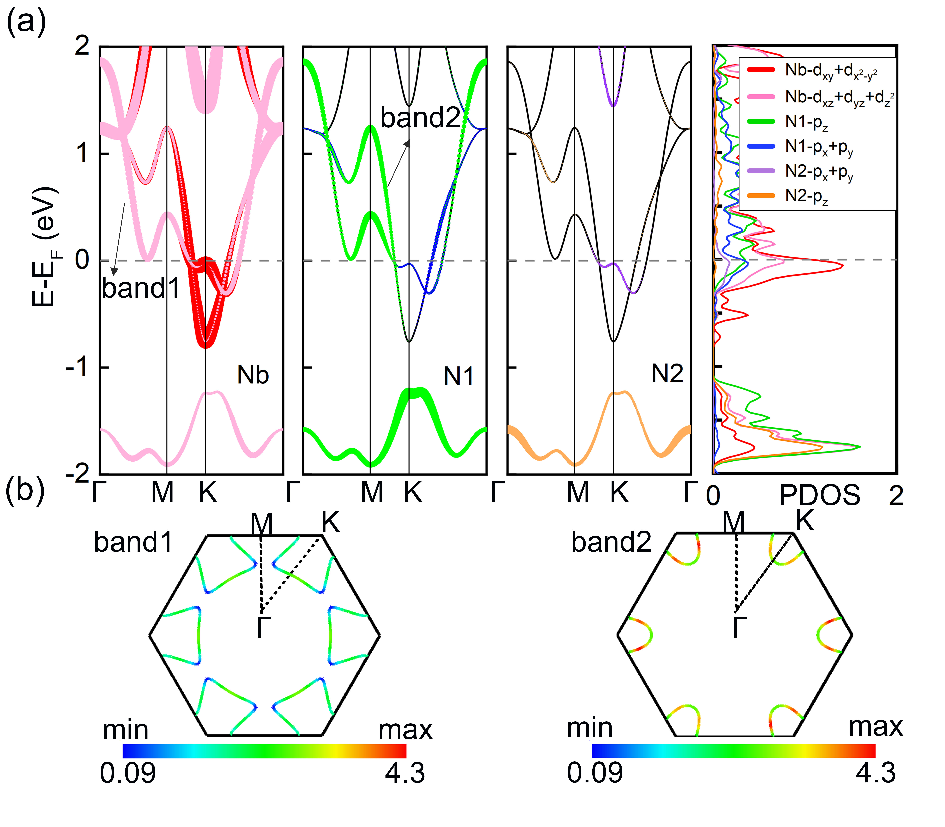}
    \caption{(a) The projected electronic band structures and PDOS of
    Nb$_2$N$_3$. (b) The band resolved FS with the relative Fermi velocity demonstrated by the color. 
    Red, green, and blue colors denote the regions that 
    have high, middle, and low Fermi velocities, respectively.}
    \label{fig2}
\end{figure}

To further understand the electronic properties of the Nb$_2$N$_3$ monolayer, the atomic orbital projected 
band structures on Nb-$d$ and N-$p$ orbitals and corresponding projected density of states (PDOS) 
are presented in Fig.~\ref{fig2}(a). There are two conduction bands labelled by band 1 and band 2 
crossing the Fermi level ($E_\mathrm{F}$), demonstrating
a electron-type semi-metallic characteristic in the Nb$_2$N$_3$ monolayer.
Figure~\ref{fig2}(b) shows the band resolved Fermi surface (FS) of the Nb$_2$N$_3$ monolayer,
in which the isovalue color indicates the relative Fermi velocity ($\nu_\mathrm{F}$). 
One can find that the ratio between the maximum and the minimum $\nu_\mathrm{F}$ is approximately 47.8, reflecting the huge difference in energy 
dispersion between band 1 and band 2. In detail,
the maximum $\nu_\mathrm{F}$ is found in the band 2 near the K point in the BZ. On the contrary, the minimum 
one occurs in the band 1, which can be attributed to 
the flat-band feature (along the M-K path) 
around the $E_\mathrm{F}$. The flat-band characteristic gives rise to a van Hove singularity, results in a DOS peak near the $E_\mathrm{F}$ [see Fig.~\ref{fig2}(b)]. 
Notably, the oscillatory behavior of the energy dispersion above $E_\mathrm{F}$ in the band 1 
further contributes to a large DOS. These characteristics make it necessary to 
explore superconducting properties of the Nb$_2$N$_3$ monolayer.
From the projected band structures and PDOS, one can see that the electronic states 
near the $E_\mathrm{F}$ are predominantly contributed by Nb-$d$ and N-$p$ orbitals. 
Specifically, the in-plane orbitals ($d_{xy}$ and $d_{x^2-y^2}$) and 
out-of-plane orbitals ($d_{xz}$, $d_{yz}$, and $d_{z^2}$) of Nb atoms make the most 
significant contributions to the bands near the $E_\mathrm{F}$, and following this are 
the $p_z$, $p_x$, and $p_y$ orbitals of the N1 atoms, with the in-plane orbitals ($p_x$ and $p_y$) 
of N2 atoms contributing the least.
It is also seen that the contributions from Nb-$d$ and N-$p$ have same and prominent DOS peaks 
around $E_\mathrm{F}$ shown in Fig.~\ref{fig2}(a), indicating a strong $p$-$d$ hybridization in the
Nb$_2$N$_3$ monolayer.
Furthermore, to elucidate the evolution of electronic properties across different 
layered structures, we have calculated the band structures and PDOS for the NbN$_2$,
Nb$_3$N$_4$, Nb$_4$N$_5$ monolayers, as presented in Fig.~S5 \cite{supp}.
Notably, the NbN$_2$ monolayer exhibits a distinct characteristic: 
the DOS at the Fermi level $N(E_\mathrm{F})$ contribution from 
N atoms exceeds that from Nb atoms, with the Nb-derived DOS being particularly small. 
In contrast, the Nb$_3$N$_4$ and Nb$_4$N$_5$ monolayers 
display similar electronic characteristics near the Fermi level, 
with the total DOS at the Fermi level situated in a local minimum or valley. 
This behavior contrasts sharply with the Nb$_2$N$_3$ monolayer, 
which exhibits a pronounced Van Hove singularity at the Fermi level, 
a feature that likely enhances its superconducting properties.

\begin{figure}[thp]
    \includegraphics[width=0.48 \textwidth]{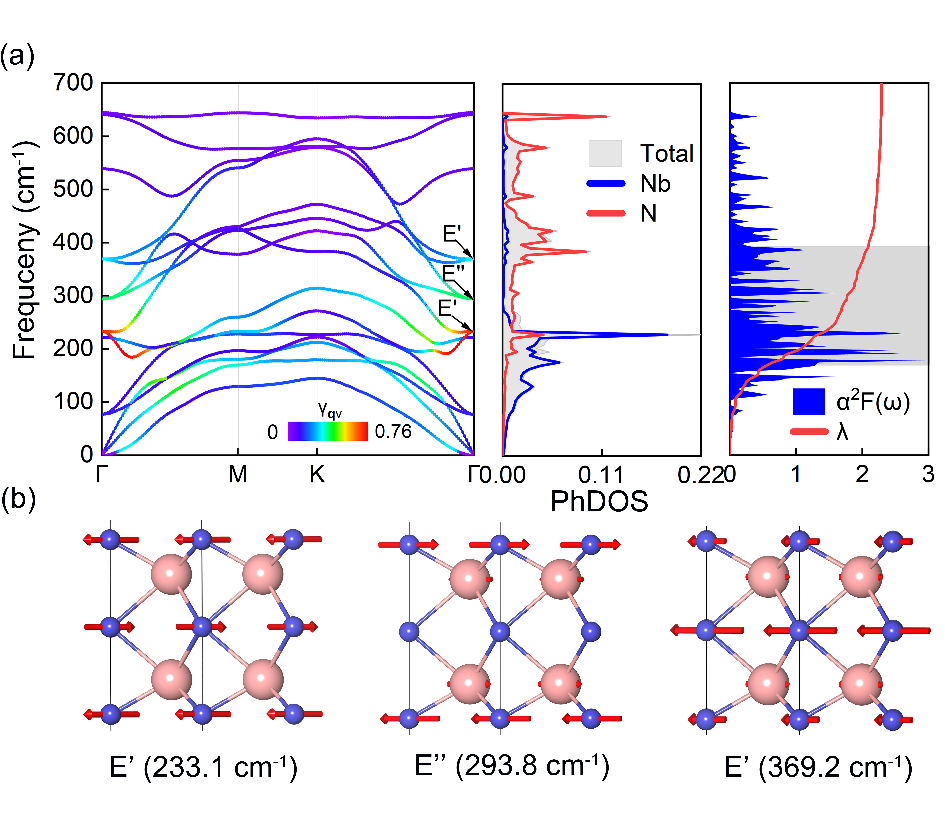}
    \caption{(a) The phonon dispersion, PhDOSs, Eliashberg spectral function $\alpha^2F(\omega)$, and integrated EPC constant     
    $\lambda (\omega)$ of the Nb$_2$N$_3$ monolayer.   
    The unit for the PhDOS is states per cm$^{-1}$ per unit cell.
   (b) Three strongly coupled vibrational patterns at $\Gamma$ point and the corresponding 
   vibration frequencies are shown below each figure. 
   Red arrows represent the atomic vibration direction .
   }\label{fig3}
\end{figure}

Then, we investigate the superconducting properties of the Nb$_2$N$_3$ monolayer, and its phonon spectra, partial atomic phonon density of states (PhDOS), Eliashberg spectral function $\alpha^2F(\omega)$, 
and the cumulative $\lambda(\omega)$ are shown in Fig.~\ref{fig3}(a). 
We find that the absence of imaginary frequencies in the phonon dispersion curves across 
the entire BZ confirms the dynamical stability of the Nb$_2$N$_3$ monolayer.
From PhDOS, the vibration modes can be separated into two regions: the low-frequency region
below 250~cm$^{-1}$ primarily originating from the vibrations of Nb atoms, and the high-frequency region
mainly contributed by N atoms,
which is consistent with expectations from the heavier mass of Nb relative to N. 
To further elucidate the nature of electron-phonon coupling, the phonon mode linewidths 
$\gamma_{\mathbf{q}\nu}$ are superimposed on the phonon dispersion curves. 
There are three largest linewidths deriving from the phonon modes at $\Gamma$ 
point: (1) $E^{\prime}$ mode at 233.1~cm$^{-1}$; 
(2) $E^{''}$ mode at 293.8~cm$^{-1}$; (3) $E^{\prime\prime}$ mode at 369.2~cm$^{-1}$.
The corresponding atomic vibrational patterns are schematically demonstrated in Fig.~\ref{fig3}(b).
Among these modes, all N atoms exhibit in-plane vibrations while the Nb atoms remain nearly stationary. 
The Nb atoms exhibit minimal movement in the lower $E^{\prime}$ phonon mode at 233.1cm$^{-1}$. 
%which contrasts with their significant contribution to the PhDOS. 
It is worth noting that the presence of soft and flat phonon modes below 250~cm$^{-1}$ along the $\Gamma$-M 
and $\Gamma$-K paths significantly contributes to the EPC strength, 
which is consistent with the $\delta$-like peaks observed in the spectral function $\alpha^2F(\omega)$.
Combining the calculated phonon linewidths with the electronic structures, we conclude that the superconductivity in Nb$_2$N$_3$ primarily arises from the strong coupling between the $E^{\prime}$ modes 
of N atoms and the electronic bands derived from N-$p_{x}$+$p_{y}$ and Nb-$d_{xy}$+$d_{x^{2}-y^{2}}$ 
states. This coupling is particularly significant for the modes at the $\Gamma$ point, 
which involve atomic displacements in the $x$-$y$ plane.
As shown in Fig.~\ref{fig3}(a), the total $\lambda$ of the Nb$_2$N$_3$ monolayer 
is found to be 2.3, which is much larger than the similar monolayer system W$_2$N$_3$ \cite{campi2021prediction,PhysRevB.103.064510}. 
One possible reason is that the strong EPC
is primarily governed by in-plane vibrational modes, a characteristic also observed in 2D LiBC \cite{PhysRevB.104.054504}, whereas in W$_2$N$_3$, it is mainly contributed by out-of-plane modes.
To ensure accuracy in the EPC constant $\lambda$, which is highly sensitive to the 
{\it $\textbf{k}$}- and {\it $\textbf{q}$}-grid sampling density used in EPC calculations, we recalculated $\alpha^2F(\omega$) and the cumulative $\lambda(\omega)$ 
using finer grids of 288$\times$288$\times$1 and 144$\times$144$\times$1 
with Wannier interpolation, and the calculated results are 
displayed in Fig.~S6 \cite{supp}. 
This refinement yields a total EPC constant $\lambda$ of about 3.7. 
We would also like to note that the chosen fine grid parameters provide well-converged results for both $\lambda$ and $\alpha^2F(\omega)$.

Next, using the McMillan-Allen-Dynes equation \cite{PhysRev.167.331,DYNES1972615,PhysRevB.12.905}, $T_\mathrm{c}=\frac{\omega_\mathrm{log}}{1.2} \exp\biggl[-\frac{1.04(1+\lambda)}{\lambda-\mu^{*}(1+0.62\lambda)}\biggr]$, 
with $\mu^{*}$ set to a typical value of 0.11 and 
a calculated $\omega_\mathrm{log}$ of 235~cm$^{-1}$, 
we estimate the $T_c$ of the Nb$_2$N$_3$ monolayer to be 47 K.
Additionally, since the calculated EPC constant $\lambda$ exceeds 2.0, 
the strong-coupling Allen-Dynes modified formula 
is more appropriate for calculating the $T_c$. The modified equation is given by
$T_\mathrm{c}=f_1f_2\frac{\omega_\mathrm{log}}{1.2} \exp\biggl[-\frac{1.04(1+\lambda)}{\lambda-\mu^{*}(1+0.62\lambda)}\biggr]$ \cite{PhysRevB.12.905}, 
where $f_1$ and $f_2$ are correction factors accounting 
for strong coupling and shape effects, respectively. 
Applying this formula with the obtained or set parameters, 
we obtained a $T_c$ of 68.9 K for the Nb$_2$N$_3$ monolayer, 
underscoring the significant role of strong EPC in 
enhancing its superconducting properties.
We have also calculated the $T_{c}$ for the other three layered phases 
using the McMillan-Allen-Dynes equation, as all 
their $\lambda$ are smaller than 1.5. 
The obtained $T_c$ values are 0.58 K for NbN$_2$, 31.4 K for 
Nb$_3$N$_4$, and 30.9 K for Nb$_4$N$_5$, 
as shown in Fig.~S2~\cite{supp}. To understand the origin of these variations in $T_c$, we analyzed the three key parameters,
namely, $\lambda$, $\omega_{\log}$, and $N(E_\mathrm{F})$.
Although the NbN$_2$ monolayer exhibits the highest $\omega_{\log}$, 
its extremely low $N(E_\mathrm{F})$ severely weakens the EPC, 
resulting in a drastically reduced $\lambda$ and consequently the lowest $T_c$ among the monolayers. For the Nb$_3$N$_4$ and Nb$_4$N$_5$ monolayers, both $N(E_\mathrm{F})$ 
and $\lambda$ are comparable. However, the slightly lower $\omega_{\log}$ of 
Nb$_4$N$_5$ leads to a marginally reduced $T_c$ compared to Nb$_3$N$_4$. 
In the Nb$_2$N$_3$ monolayer, although $\omega_{\log}$ decreases to approximately 310~K, 
the significantly enhanced $N(E_\mathrm{F})$ and $\lambda$ result in the highest $T_c$.
From these results, it is indicated that the influence of $N(E_F)$ and $\lambda$ 
on the $T_c$ is more pronounced than that of $\omega_{\log}$ for our predicted Nb-N nanosheets.
Meanwhile, it is important to know the impact of $\mu^*$ on $T_c$, as the predicted $T_c$ value is sensitive to the choice of $\mu^*$. 
To address this, we systematically investigated the $T_c$ of the Nb$_2$N$_3$ monolayer as a function of $\mu^*$, with $\mu^*$ values ranging from 0.1 to 0.13. 
As shown in Table~S2~\cite{supp}, the $T_c$ of the Nb$_2$N$_3$ monolayer decreases monotonically with increasing $\mu^*$, from 70.6~K at $\mu^*$ = 0.1 to 65.4~K at $\mu^*$ = 0.13. Notably, for $\mu^*$ = 0.13, 
the predicted $T_c$ remains significantly elevated (65.4~K), 
demonstrating remarkable resilience against Coulomb pseudopotential variations. 
This robustness stems from the inherently strong EPC in the Nb$_2$N$_3$ system, which dominates over the $\mu^*$-induced suppression. 
Crucially, all calculated $T_c$ values substantially exceed the conventional superconducting regime thereby preserving the central conclusion of high-temperature superconductivity in this material.

\begin{figure}[thp]
	\includegraphics[width=0.5 \textwidth]{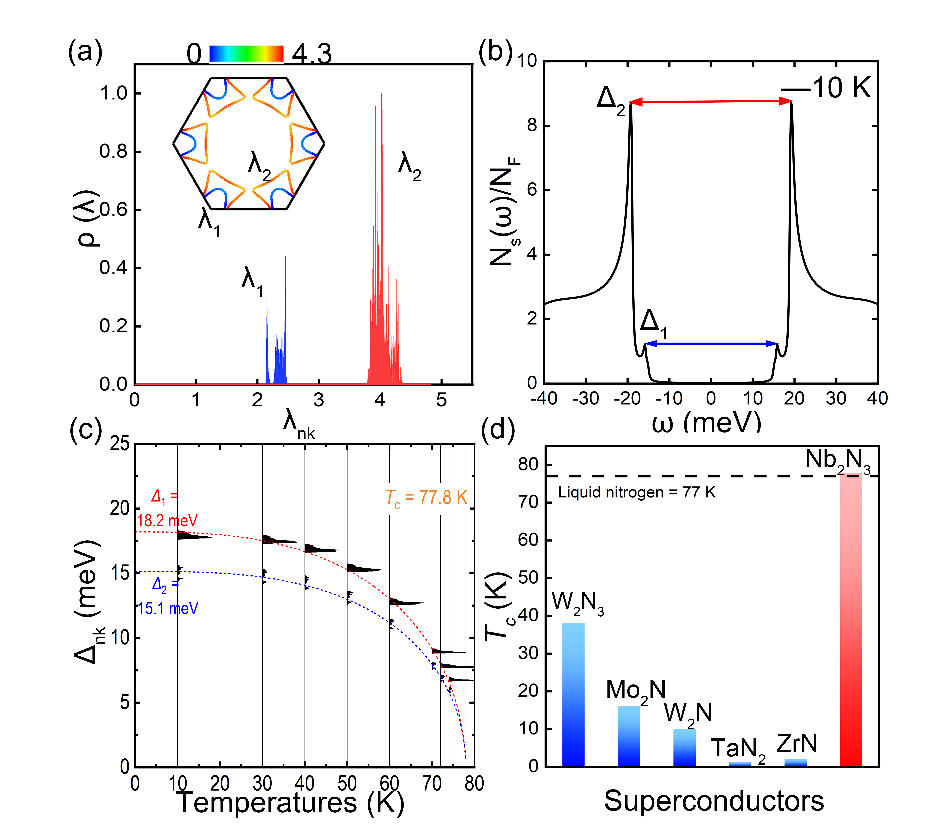}
	\caption{(a) Distribution $\rho$($\lambda$) of the {\it \textbf{k}}-resolved EPC constant $\lambda_{nk}$ associated with the inset of the $\lambda_{nk}$ projected on to each FS sheet. The two separated regimes of $\lambda_{nk}$ are indicated as $\lambda_1$ and $\lambda_2$.  
	(b) Calculated normalized quasi-particle DOS in the superconducting state 
	of the Nb$_2$N$_3$ monolayer. 
	(c) Temperature-dependent superconducting gap distribution of the Nb$_2$N$_3$ monolayer. 
	%The dashed line in c represents $\alpha$ model fitting using the average gap values. 
	(d) The $T_c$ values for some typical superconducting TMNs~\cite{D0NR03875J,guo2019freestanding,campi2021prediction,PhysRevB.103.064510}.
}\label{fig4}
\end{figure}

%The superconducting energy gap $\Delta_{n\boldsymbol{k}}$ and the EPC parameter $\lambda_{n\boldsymbol{k}}$ are .

Due to the reduced dimensionality in monolayer systems and the inherent 
anisotropy of the EPC, the isotropic approximation applied in the Allen-Dynes formula 
may not accurately capture the EPC effects and superconducting gap 
characteristics \cite{PhysRevB.103.064510}. 
To investigate the impact of this anisotropy on the superconducting properties, 
we further numerically solve the fully anisotropic 
Migdal-Eliashberg equations based on electron-phonon Wannier-Fourier interpolation,
which is implemented in EPW code~\cite{PhysRevB.76.165108,PhysRevB.87.024505,ponce2016epw}.
As illustrated in Fig.~S7 \cite{supp},
the maximally localized Wannier function (MLWF) interpolation accurately 
reproduces the band structure obtained from DFT calculations,  thereby providing a reliable basis for the subsequent EPW calculations.
To quantify the anisotropy of the EPC \cite{PhysRevMaterials.7.114805}, we calculated 
the corresponding projections of the EPC strength $\lambda_{nk}$ on each FS sheets, 
as depicted in Fig.~\ref{fig4}(a).
It is observed that the distribution of $\lambda_{nk}$ on the FS is significantly anisotropic, featuring two distinct regimes: a smaller regime $\lambda_{1}$ in the range of 2.1-2.5, associated with the FS2 
sheets derived from $d_{xy}$ and $d_{x^2-y^2}$ orbitals of Nb atoms, 
and a larger regime $\lambda_{2}$ in the range of 4.0-4.3, also linked to the FS1 sheets of Nb $d_{xy}$+$d_{x^2-y^2}$ orbitals.
This behavior contrasts with the multi-gap superconductors MgB$_2$~\cite{PhysRevB.66.020513} 
and LiBC \cite{PhysRevB.104.054504}, 
where the FS arise from different orbitals, leading to anisotropic distributions.
The main reason for this anisotropy is that the electrons around the Fermi level are 
predominantly contributed by FS1, resulting in a larger EPC on this sheet.

Then, the normalized quasiparticle DOS at 10~K for the Nb$_2$N$_3$ monolayer 
is plotted in Fig.~\ref{fig4}(b). It is clear that
Nb$_2$N$_3$ exhibits a highly anisotropic two-gap characteristic, 
as evidenced by the superconducting DOS displaying two peaks at 16~meV and 19~meV, respectively,
which coincide with the calculated superconducting gaps shown in Fig.~\ref{fig4}(c).
The anisotropic Eliashberg equations at different temperatures 
are then solved to further determine the superconducting gaps $\Delta_{nk}$ [see Fig.~\ref{fig4}(c)]. 
It is found that two $\Delta_{nk}$ are clearly observed and both of
them vanishes at a critical temperature $T_c$ of about 77.8~K, 
which is significantly greater than those (68.9~K) obtained from the strong coupling Allen-Dynes formula.
This highlights the significance of anisotropic effects in 
characterizing the superconducting properties of the Nb$_2$N$_3$ monolayer.
At the zero-temperature limit, 
the large superconducting gap of the Nb$_2$N$_3$ monolayer is estimated to be $\Delta_1$ = 18.2~meV, 
indicating significant Cooper pairing with the ratio of 2$\Delta_1$/$k_\mathrm{B}T_c$ = 5.4 $\textgreater$ 3.5, 
while the smaller gap $\Delta_2$ =15.1~meV has the ratio of 2$\Delta_2$/$k_\mathrm{B}T_c$ = 5.4 $\textgreater$ 3.5 \cite{PhysRevMaterials.7.114805}.
By fitting the temperature dependence using $\Delta(T)= \Delta(T=0)[(1-(T/T_c))^p]^{0.5}$, 
where $\Delta(T=0)$ is the gap at zero temperature and $p$ is fitting exponent, 
we obtained $p_1$ = 1.8 and $p_2$ = 2.0. We note that $p$ is greater than 1.5, 
indicating that the superconducting gap of the superconductor is relatively large. Moreover, near the $T_c$, the superconducting gap is extremely sensitive to temperature changes. Therefore, Nb$_2$N$_3$ may be used as a superconducting switch device~\cite{PhysRevLett.94.157005,CLEM1966268}.
Furthermore, we compared the $T_c$ values of previously reported 2D binary superconductors, particularly TMNs, as shown in Fig.~\ref{fig4}(d). 
Notably, the Nb$_2$N$_3$ monolayer exhibits the highest $T_c$ of 77.8~K among these materials. This significant $T_c$, which exceeds the liquid nitrogen temperature (77~K) and sets a new record for the highest $T_c$ in 2D binary superconductors, suggesting that 2D Nb$_2$N$_3$ superconductor could be promising candidates for future applications, especially in the field of nano-superconductivity.

\section{Conclusion}
In summary, we report a series of 2D Nb$_x$N$_{x+1}$ ($x$=1-4) monolayers obtained from cleaving bulk NbN, investigated through first-principles calculations. 
Four monolayers: NbN$_2$, Nb$_2$N$_3$, Nb$_3$N$_4$, and Nb$_4$N$_5$, are found to be dynamically stable. It is found that these monolayers exhibit previously 
unexplored phonon-mediated superconductivity, with Nb$_2$N$_3$ predicted 
to have the highest $T_c$. Detailed analysis of the Nb$_2$N$_3$ monolayer 
reveals that phonons associated with N atoms, coupled with electrons 
from N-$p$ and Nb-$d$ orbitals, contribute to a strong EPC, crucial for 
achieving high-temperature superconductivity. 
The fully anisotropic Eliashberg theory reveals that the Nb$_2$N$_3$ 
monolayer exhibits a pronounced EPC anisotropy and a distinctive two-gap superconducting state, 
which results in Nb$_2$N$_3$ achieving a record high $T_c$ of 77.8 K among binary superconductors.
These findings provide new insights into the superconducting 
mechanisms of transition metal nitride monolayers and open avenues 
for applications in nanoscale superconducting devices.

\section{acknowledgments}
We acknowledge support from the National Natural Science Foundation of China (Grants No.~12074153 and ~12204202), the Natural Science Foundation of Jiangsu Province (Grant No.~BK20220679), 
and the Natural Science Fund for Colleges and Universities in Jiangsu 
Province (Grant No.~22KJB140010).

\bibliography{ref}

\begin{thebibliography}{67}%
\makeatletter
\providecommand \@ifxundefined [1]{%
 \@ifx{#1\undefined}
}%
\providecommand \@ifnum [1]{%
 \ifnum #1\expandafter \@firstoftwo
 \else \expandafter \@secondoftwo
 \fi
}%
\providecommand \@ifx [1]{%
 \ifx #1\expandafter \@firstoftwo
 \else \expandafter \@secondoftwo
 \fi
}%
\providecommand \natexlab [1]{#1}%
\providecommand \enquote  [1]{``#1''}%
\providecommand \bibnamefont  [1]{#1}%
\providecommand \bibfnamefont [1]{#1}%
\providecommand \citenamefont [1]{#1}%
\providecommand \href@noop [0]{\@secondoftwo}%
\providecommand \href [0]{\begingroup \@sanitize@url \@href}%
\providecommand \@href[1]{\@@startlink{#1}\@@href}%
\providecommand \@@href[1]{\endgroup#1\@@endlink}%
\providecommand \@sanitize@url [0]{\catcode `\\12\catcode `\$12\catcode
  `\&12\catcode `\#12\catcode `\^12\catcode `\_12\catcode `\%12\relax}%
\providecommand \@@startlink[1]{}%
\providecommand \@@endlink[0]{}%
\providecommand \url  [0]{\begingroup\@sanitize@url \@url }%
\providecommand \@url [1]{\endgroup\@href {#1}{\urlprefix }}%
\providecommand \urlprefix  [0]{URL }%
\providecommand \Eprint [0]{\href }%
\providecommand \doibase [0]{https://doi.org/}%
\providecommand \selectlanguage [0]{\@gobble}%
\providecommand \bibinfo  [0]{\@secondoftwo}%
\providecommand \bibfield  [0]{\@secondoftwo}%
\providecommand \translation [1]{[#1]}%
\providecommand \BibitemOpen [0]{}%
\providecommand \bibitemStop [0]{}%
\providecommand \bibitemNoStop [0]{.\EOS\space}%
\providecommand \EOS [0]{\spacefactor3000\relax}%
\providecommand \BibitemShut  [1]{\csname bibitem#1\endcsname}%
\let\auto@bib@innerbib\@empty
%</preamble>
\bibitem [{\citenamefont {Lu}\ \emph {et~al.}(2015)\citenamefont {Lu},
  \citenamefont {Zheliuk}, \citenamefont {Leermakers}, \citenamefont {Yuan},
  \citenamefont {Zeitler}, \citenamefont {Law},\ and\ \citenamefont
  {Ye}}]{doi:10.1126/science.aab2277}%
  \BibitemOpen
  \bibfield  {author} {\bibinfo {author} {\bibfnamefont {J.~M.}\ \bibnamefont
  {Lu}}, \bibinfo {author} {\bibfnamefont {O.}~\bibnamefont {Zheliuk}},
  \bibinfo {author} {\bibfnamefont {I.}~\bibnamefont {Leermakers}}, \bibinfo
  {author} {\bibfnamefont {N.~F.~Q.}\ \bibnamefont {Yuan}}, \bibinfo {author}
  {\bibfnamefont {U.}~\bibnamefont {Zeitler}}, \bibinfo {author} {\bibfnamefont
  {K.~T.}\ \bibnamefont {Law}},\ and\ \bibinfo {author} {\bibfnamefont {J.~T.}\
  \bibnamefont {Ye}},\ }\bibfield  {title} {\bibinfo {title} {{Evidence for
  two-dimensional Ising superconductivity in gated MoS$_2$}},\ }\href
  {https://doi.org/10.1126/science.aab2277} {\bibfield  {journal} {\bibinfo
  {journal} {Science}\ }\textbf {\bibinfo {volume} {350}},\ \bibinfo {pages}
  {1353} (\bibinfo {year} {2015})}\BibitemShut {NoStop}%
\bibitem [{\citenamefont {Xi}\ \emph {et~al.}(2016)\citenamefont {Xi},
  \citenamefont {Wang}, \citenamefont {Zhao}, \citenamefont {Park},
  \citenamefont {Law}, \citenamefont {Berger}, \citenamefont {Forr{\'o}},
  \citenamefont {Shan},\ and\ \citenamefont {Mak}}]{xi2016ising}%
  \BibitemOpen
  \bibfield  {author} {\bibinfo {author} {\bibfnamefont {X.}~\bibnamefont
  {Xi}}, \bibinfo {author} {\bibfnamefont {Z.}~\bibnamefont {Wang}}, \bibinfo
  {author} {\bibfnamefont {W.}~\bibnamefont {Zhao}}, \bibinfo {author}
  {\bibfnamefont {J.-H.}\ \bibnamefont {Park}}, \bibinfo {author}
  {\bibfnamefont {K.~T.}\ \bibnamefont {Law}}, \bibinfo {author} {\bibfnamefont
  {H.}~\bibnamefont {Berger}}, \bibinfo {author} {\bibfnamefont
  {L.}~\bibnamefont {Forr{\'o}}}, \bibinfo {author} {\bibfnamefont
  {J.}~\bibnamefont {Shan}},\ and\ \bibinfo {author} {\bibfnamefont {K.~F.}\
  \bibnamefont {Mak}},\ }\bibfield  {title} {\bibinfo {title} {{Ising pairing
  in superconducting NbSe$_2$ atomic layers}},\ }\href
  {https://doi.org/10.1038/nphys3538} {\bibfield  {journal} {\bibinfo
  {journal} {Nat. Phys.}\ }\textbf {\bibinfo {volume} {12}},\ \bibinfo {pages}
  {139} (\bibinfo {year} {2016})}\BibitemShut {NoStop}%
\bibitem [{\citenamefont {Qiu}\ \emph {et~al.}(2021)\citenamefont {Qiu},
  \citenamefont {Gong}, \citenamefont {Wang}, \citenamefont {Zhang},
  \citenamefont {Yang}, \citenamefont {Wang},\ and\ \citenamefont
  {Xiong}}]{hadma.202006124}%
  \BibitemOpen
  \bibfield  {author} {\bibinfo {author} {\bibfnamefont {D.}~\bibnamefont
  {Qiu}}, \bibinfo {author} {\bibfnamefont {C.}~\bibnamefont {Gong}}, \bibinfo
  {author} {\bibfnamefont {S.}~\bibnamefont {Wang}}, \bibinfo {author}
  {\bibfnamefont {M.}~\bibnamefont {Zhang}}, \bibinfo {author} {\bibfnamefont
  {C.}~\bibnamefont {Yang}}, \bibinfo {author} {\bibfnamefont {X.}~\bibnamefont
  {Wang}},\ and\ \bibinfo {author} {\bibfnamefont {J.}~\bibnamefont {Xiong}},\
  }\bibfield  {title} {\bibinfo {title} {{Recent Advances in 2D
  Superconductors}},\ }\href
  {https://doi.org/https://doi.org/10.1002/adma.202006124} {\bibfield
  {journal} {\bibinfo  {journal} {Adv. Mater.}\ }\textbf {\bibinfo {volume}
  {33}},\ \bibinfo {pages} {2006124} (\bibinfo {year} {2021})}\BibitemShut
  {NoStop}%
\bibitem [{\citenamefont {Lu}\ \emph {et~al.}(2024)\citenamefont {Lu},
  \citenamefont {Han}, \citenamefont {Yao}, \citenamefont {Reddy},
  \citenamefont {Yang}, \citenamefont {Seo}, \citenamefont {Watanabe},
  \citenamefont {Taniguchi}, \citenamefont {Fu},\ and\ \citenamefont
  {Ju}}]{lu2024fractional}%
  \BibitemOpen
  \bibfield  {author} {\bibinfo {author} {\bibfnamefont {Z.}~\bibnamefont
  {Lu}}, \bibinfo {author} {\bibfnamefont {T.}~\bibnamefont {Han}}, \bibinfo
  {author} {\bibfnamefont {Y.}~\bibnamefont {Yao}}, \bibinfo {author}
  {\bibfnamefont {A.~P.}\ \bibnamefont {Reddy}}, \bibinfo {author}
  {\bibfnamefont {J.}~\bibnamefont {Yang}}, \bibinfo {author} {\bibfnamefont
  {J.}~\bibnamefont {Seo}}, \bibinfo {author} {\bibfnamefont {K.}~\bibnamefont
  {Watanabe}}, \bibinfo {author} {\bibfnamefont {T.}~\bibnamefont {Taniguchi}},
  \bibinfo {author} {\bibfnamefont {L.}~\bibnamefont {Fu}},\ and\ \bibinfo
  {author} {\bibfnamefont {L.}~\bibnamefont {Ju}},\ }\bibfield  {title}
  {\bibinfo {title} {{Fractional quantum anomalous Hall effect in multilayer
  graphene}},\ }\href {https://doi.org/0.1038/s41586-023-07010-7} {\bibfield
  {journal} {\bibinfo  {journal} {Nature}\ }\textbf {\bibinfo {volume} {626}},\
  \bibinfo {pages} {759} (\bibinfo {year} {2024})}\BibitemShut {NoStop}%
\bibitem [{\citenamefont {Jiang}\ \emph {et~al.}(2024)\citenamefont {Jiang},
  \citenamefont {Wang}, \citenamefont {Bao}, \citenamefont {Liu},\ and\
  \citenamefont {Wang}}]{PhysRevLett.132.106602}%
  \BibitemOpen
  \bibfield  {author} {\bibinfo {author} {\bibfnamefont {Y.}~\bibnamefont
  {Jiang}}, \bibinfo {author} {\bibfnamefont {H.}~\bibnamefont {Wang}},
  \bibinfo {author} {\bibfnamefont {K.}~\bibnamefont {Bao}}, \bibinfo {author}
  {\bibfnamefont {Z.}~\bibnamefont {Liu}},\ and\ \bibinfo {author}
  {\bibfnamefont {J.}~\bibnamefont {Wang}},\ }\bibfield  {title} {\bibinfo
  {title} {{Monolayer ${\mathrm{V}}_{2}{MX}_{4}$: A New Family of Quantum
  Anomalous Hall Insulators}},\ }\href
  {https://doi.org/10.1103/PhysRevLett.132.106602} {\bibfield  {journal}
  {\bibinfo  {journal} {Phys. Rev. Lett.}\ }\textbf {\bibinfo {volume} {132}},\
  \bibinfo {pages} {106602} (\bibinfo {year} {2024})}\BibitemShut {NoStop}%
\bibitem [{\citenamefont {Tong}\ \emph {et~al.}(2016)\citenamefont {Tong},
  \citenamefont {Gong}, \citenamefont {Wan},\ and\ \citenamefont
  {Duan}}]{tong2016concepts}%
  \BibitemOpen
  \bibfield  {author} {\bibinfo {author} {\bibfnamefont {W.-Y.}\ \bibnamefont
  {Tong}}, \bibinfo {author} {\bibfnamefont {S.-J.}\ \bibnamefont {Gong}},
  \bibinfo {author} {\bibfnamefont {X.}~\bibnamefont {Wan}},\ and\ \bibinfo
  {author} {\bibfnamefont {C.-G.}\ \bibnamefont {Duan}},\ }\bibfield  {title}
  {\bibinfo {title} {{Concepts of ferrovalley material and anomalous valley
  Hall effect}z},\ }\href {https://doi.org/10.1038/ncomms13612} {\bibfield
  {journal} {\bibinfo  {journal} {Nat. Commun.}\ }\textbf {\bibinfo {volume}
  {7}},\ \bibinfo {pages} {1} (\bibinfo {year} {2016})}\BibitemShut {NoStop}%
\bibitem [{\citenamefont {Jiang}\ \emph
  {et~al.}(2021{\natexlab{a}})\citenamefont {Jiang}, \citenamefont {Kang},
  \citenamefont {Li}, \citenamefont {Zheng}, \citenamefont {Zeng},\ and\
  \citenamefont {Sanvito}}]{PhysRevB.104.035430}%
  \BibitemOpen
  \bibfield  {author} {\bibinfo {author} {\bibfnamefont {P.}~\bibnamefont
  {Jiang}}, \bibinfo {author} {\bibfnamefont {L.}~\bibnamefont {Kang}},
  \bibinfo {author} {\bibfnamefont {Y.-L.}\ \bibnamefont {Li}}, \bibinfo
  {author} {\bibfnamefont {X.}~\bibnamefont {Zheng}}, \bibinfo {author}
  {\bibfnamefont {Z.}~\bibnamefont {Zeng}},\ and\ \bibinfo {author}
  {\bibfnamefont {S.}~\bibnamefont {Sanvito}},\ }\bibfield  {title} {\bibinfo
  {title} {{Prediction of the two-dimensional Janus ferrovalley material
  LaBrI}},\ }\href {https://doi.org/10.1103/PhysRevB.104.035430} {\bibfield
  {journal} {\bibinfo  {journal} {Phys. Rev. B}\ }\textbf {\bibinfo {volume}
  {104}},\ \bibinfo {pages} {035430} (\bibinfo {year}
  {2021}{\natexlab{a}})}\BibitemShut {NoStop}%
\bibitem [{\citenamefont {Jiang}\ \emph
  {et~al.}(2021{\natexlab{b}})\citenamefont {Jiang}, \citenamefont {Tao},
  \citenamefont {Hao}, \citenamefont {Liu},\ and\ \citenamefont
  {Zeng}}]{2021Two}%
  \BibitemOpen
  \bibfield  {author} {\bibinfo {author} {\bibfnamefont {P.}~\bibnamefont
  {Jiang}}, \bibinfo {author} {\bibfnamefont {X.}~\bibnamefont {Tao}}, \bibinfo
  {author} {\bibfnamefont {H.}~\bibnamefont {Hao}}, \bibinfo {author}
  {\bibfnamefont {Y.}~\bibnamefont {Liu}},\ and\ \bibinfo {author}
  {\bibfnamefont {Z.}~\bibnamefont {Zeng}},\ }\bibfield  {title} {\bibinfo
  {title} {{Two-dimensional centrosymmetrical antiferromagnets for spin
  photogalvanic devices}},\ }\href {https://doi.org/10.1038/s41534-021-00365-7}
  {\bibfield  {journal} {\bibinfo  {journal} {npj Quantum Inf.}\ }\textbf
  {\bibinfo {volume} {7}},\ \bibinfo {pages} {21} (\bibinfo {year}
  {2021}{\natexlab{b}})}\BibitemShut {NoStop}%
\bibitem [{\citenamefont {Tao}\ \emph {et~al.}(2020)\citenamefont {Tao},
  \citenamefont {Jiang}, \citenamefont {Hao}, \citenamefont {Zheng},
  \citenamefont {Zhang},\ and\ \citenamefont {Zeng}}]{PhysRevB.102.081402}%
  \BibitemOpen
  \bibfield  {author} {\bibinfo {author} {\bibfnamefont {X.}~\bibnamefont
  {Tao}}, \bibinfo {author} {\bibfnamefont {P.}~\bibnamefont {Jiang}}, \bibinfo
  {author} {\bibfnamefont {H.}~\bibnamefont {Hao}}, \bibinfo {author}
  {\bibfnamefont {X.}~\bibnamefont {Zheng}}, \bibinfo {author} {\bibfnamefont
  {L.}~\bibnamefont {Zhang}},\ and\ \bibinfo {author} {\bibfnamefont
  {Z.}~\bibnamefont {Zeng}},\ }\bibfield  {title} {\bibinfo {title} {Pure spin
  current generation via photogalvanic effect with spatial inversion
  symmetry},\ }\href {https://doi.org/10.1103/PhysRevB.102.081402} {\bibfield
  {journal} {\bibinfo  {journal} {Phys. Rev. B}\ }\textbf {\bibinfo {volume}
  {102}},\ \bibinfo {pages} {081402} (\bibinfo {year} {2020})}\BibitemShut
  {NoStop}%
\bibitem [{\citenamefont {Xu}\ \emph {et~al.}(2020)\citenamefont {Xu},
  \citenamefont {Feng}, \citenamefont {Prokhorenko}, \citenamefont {Nahas},
  \citenamefont {Xiang},\ and\ \citenamefont
  {Bellaiche}}]{PhysRevB.101.060404}%
  \BibitemOpen
  \bibfield  {author} {\bibinfo {author} {\bibfnamefont {C.}~\bibnamefont
  {Xu}}, \bibinfo {author} {\bibfnamefont {J.}~\bibnamefont {Feng}}, \bibinfo
  {author} {\bibfnamefont {S.}~\bibnamefont {Prokhorenko}}, \bibinfo {author}
  {\bibfnamefont {Y.}~\bibnamefont {Nahas}}, \bibinfo {author} {\bibfnamefont
  {H.}~\bibnamefont {Xiang}},\ and\ \bibinfo {author} {\bibfnamefont
  {L.}~\bibnamefont {Bellaiche}},\ }\bibfield  {title} {\bibinfo {title}
  {{Topological spin texture in Janus monolayers of the chromium trihalides
  Cr(I, X$_3$)}},\ }\href {https://doi.org/10.1103/PhysRevB.101.060404}
  {\bibfield  {journal} {\bibinfo  {journal} {Phys. Rev. B}\ }\textbf {\bibinfo
  {volume} {101}},\ \bibinfo {pages} {060404(R)} (\bibinfo {year}
  {2020})}\BibitemShut {NoStop}%
\bibitem [{\citenamefont {Hsu}\ \emph {et~al.}(2017)\citenamefont {Hsu},
  \citenamefont {Vaezi}, \citenamefont {Fischer},\ and\ \citenamefont
  {Kim}}]{hsu2017topological}%
  \BibitemOpen
  \bibfield  {author} {\bibinfo {author} {\bibfnamefont {Y.-T.}\ \bibnamefont
  {Hsu}}, \bibinfo {author} {\bibfnamefont {A.}~\bibnamefont {Vaezi}}, \bibinfo
  {author} {\bibfnamefont {M.~H.}\ \bibnamefont {Fischer}},\ and\ \bibinfo
  {author} {\bibfnamefont {E.-A.}\ \bibnamefont {Kim}},\ }\bibfield  {title}
  {\bibinfo {title} {{Topological superconductivity in monolayer transition
  metal dichalcogenides}},\ }\href {https://doi.org/10.1038/ncomms14985}
  {\bibfield  {journal} {\bibinfo  {journal} {Nat. Commun.}\ }\textbf {\bibinfo
  {volume} {8}},\ \bibinfo {pages} {1} (\bibinfo {year} {2017})}\BibitemShut
  {NoStop}%
\bibitem [{\citenamefont {You}\ \emph {et~al.}(2021)\citenamefont {You},
  \citenamefont {Gu}, \citenamefont {Su},\ and\ \citenamefont
  {Feng}}]{PhysRevB.103.104503}%
  \BibitemOpen
  \bibfield  {author} {\bibinfo {author} {\bibfnamefont {J.-Y.}\ \bibnamefont
  {You}}, \bibinfo {author} {\bibfnamefont {B.}~\bibnamefont {Gu}}, \bibinfo
  {author} {\bibfnamefont {G.}~\bibnamefont {Su}},\ and\ \bibinfo {author}
  {\bibfnamefont {Y.~P.}\ \bibnamefont {Feng}},\ }\bibfield  {title} {\bibinfo
  {title} {{Two-dimensional topological superconductivity candidate in a van
  der Waals layered material}},\ }\href
  {https://doi.org/10.1103/PhysRevB.103.104503} {\bibfield  {journal} {\bibinfo
   {journal} {Phys. Rev. B}\ }\textbf {\bibinfo {volume} {103}},\ \bibinfo
  {pages} {104503} (\bibinfo {year} {2021})}\BibitemShut {NoStop}%
\bibitem [{\citenamefont {Yang}\ \emph {et~al.}(2019)\citenamefont {Yang},
  \citenamefont {Liu}, \citenamefont {Wang}, \citenamefont {Feng},
  \citenamefont {He}, \citenamefont {Sun}, \citenamefont {Tang}, \citenamefont
  {Wu}, \citenamefont {Xiong}, \citenamefont {Zhang}, \citenamefont {Lin},
  \citenamefont {Yao}, \citenamefont {Liu}, \citenamefont {Fernandes},
  \citenamefont {Xu}, \citenamefont {Valles}, \citenamefont {Wang},\ and\
  \citenamefont {Li}}]{doi:10.1126/science.aax5798}%
  \BibitemOpen
  \bibfield  {author} {\bibinfo {author} {\bibfnamefont {C.}~\bibnamefont
  {Yang}}, \bibinfo {author} {\bibfnamefont {Y.}~\bibnamefont {Liu}}, \bibinfo
  {author} {\bibfnamefont {Y.}~\bibnamefont {Wang}}, \bibinfo {author}
  {\bibfnamefont {L.}~\bibnamefont {Feng}}, \bibinfo {author} {\bibfnamefont
  {Q.}~\bibnamefont {He}}, \bibinfo {author} {\bibfnamefont {J.}~\bibnamefont
  {Sun}}, \bibinfo {author} {\bibfnamefont {Y.}~\bibnamefont {Tang}}, \bibinfo
  {author} {\bibfnamefont {C.}~\bibnamefont {Wu}}, \bibinfo {author}
  {\bibfnamefont {J.}~\bibnamefont {Xiong}}, \bibinfo {author} {\bibfnamefont
  {W.}~\bibnamefont {Zhang}}, \bibinfo {author} {\bibfnamefont
  {X.}~\bibnamefont {Lin}}, \bibinfo {author} {\bibfnamefont {H.}~\bibnamefont
  {Yao}}, \bibinfo {author} {\bibfnamefont {H.}~\bibnamefont {Liu}}, \bibinfo
  {author} {\bibfnamefont {G.}~\bibnamefont {Fernandes}}, \bibinfo {author}
  {\bibfnamefont {J.}~\bibnamefont {Xu}}, \bibinfo {author} {\bibfnamefont
  {J.~M.}\ \bibnamefont {Valles}}, \bibinfo {author} {\bibfnamefont
  {J.}~\bibnamefont {Wang}},\ and\ \bibinfo {author} {\bibfnamefont
  {Y.}~\bibnamefont {Li}},\ }\bibfield  {title} {\bibinfo {title} {Intermediate
  bosonic metallic state in the superconductor-insulator transition},\ }\href
  {https://doi.org/10.1126/science.aax5798} {\bibfield  {journal} {\bibinfo
  {journal} {Science}\ }\textbf {\bibinfo {volume} {366}},\ \bibinfo {pages}
  {1505} (\bibinfo {year} {2019})}\BibitemShut {NoStop}%
\bibitem [{\citenamefont {Liu}\ \emph {et~al.}(2021)\citenamefont {Liu},
  \citenamefont {Qi}, \citenamefont {Fang}, \citenamefont {Sun}, \citenamefont
  {Liu}, \citenamefont {Liu}, \citenamefont {Qi}, \citenamefont {Xing},
  \citenamefont {Liu}, \citenamefont {Lin}, \citenamefont {Wang}, \citenamefont
  {Xue}, \citenamefont {Xie},\ and\ \citenamefont
  {Wang}}]{PhysRevLett.127.137001}%
  \BibitemOpen
  \bibfield  {author} {\bibinfo {author} {\bibfnamefont {Y.}~\bibnamefont
  {Liu}}, \bibinfo {author} {\bibfnamefont {S.}~\bibnamefont {Qi}}, \bibinfo
  {author} {\bibfnamefont {J.}~\bibnamefont {Fang}}, \bibinfo {author}
  {\bibfnamefont {J.}~\bibnamefont {Sun}}, \bibinfo {author} {\bibfnamefont
  {C.}~\bibnamefont {Liu}}, \bibinfo {author} {\bibfnamefont {Y.}~\bibnamefont
  {Liu}}, \bibinfo {author} {\bibfnamefont {J.}~\bibnamefont {Qi}}, \bibinfo
  {author} {\bibfnamefont {Y.}~\bibnamefont {Xing}}, \bibinfo {author}
  {\bibfnamefont {H.}~\bibnamefont {Liu}}, \bibinfo {author} {\bibfnamefont
  {X.}~\bibnamefont {Lin}}, \bibinfo {author} {\bibfnamefont {L.}~\bibnamefont
  {Wang}}, \bibinfo {author} {\bibfnamefont {Q.-K.}\ \bibnamefont {Xue}},
  \bibinfo {author} {\bibfnamefont {X.~C.}\ \bibnamefont {Xie}},\ and\ \bibinfo
  {author} {\bibfnamefont {J.}~\bibnamefont {Wang}},\ }\bibfield  {title}
  {\bibinfo {title} {Observation of in-plane quantum griffiths singularity in
  two-dimensional crystalline superconductors},\ }\href
  {https://doi.org/10.1103/PhysRevLett.127.137001} {\bibfield  {journal}
  {\bibinfo  {journal} {Phys. Rev. Lett.}\ }\textbf {\bibinfo {volume} {127}},\
  \bibinfo {pages} {137001} (\bibinfo {year} {2021})}\BibitemShut {NoStop}%
\bibitem [{\citenamefont {Saito}\ \emph {et~al.}(2016)\citenamefont {Saito},
  \citenamefont {Nojima},\ and\ \citenamefont {Iwasa}}]{saito2016highly}%
  \BibitemOpen
  \bibfield  {author} {\bibinfo {author} {\bibfnamefont {Y.}~\bibnamefont
  {Saito}}, \bibinfo {author} {\bibfnamefont {T.}~\bibnamefont {Nojima}},\ and\
  \bibinfo {author} {\bibfnamefont {Y.}~\bibnamefont {Iwasa}},\ }\bibfield
  {title} {\bibinfo {title} {{Highly crystalline 2D superconductors}},\ }\href
  {https://doi.org/10.1038/natrevmats.2016.94} {\bibfield  {journal} {\bibinfo
  {journal} {Nat. Rev. Mater}\ }\textbf {\bibinfo {volume} {2}},\ \bibinfo
  {pages} {1} (\bibinfo {year} {2016})}\BibitemShut {NoStop}%
\bibitem [{\citenamefont {Bekaert}\ \emph {et~al.}(2017)\citenamefont
  {Bekaert}, \citenamefont {Aperis}, \citenamefont {Partoens}, \citenamefont
  {Oppeneer},\ and\ \citenamefont {Milo\ifmmode \check{s}\else
  \v{s}\fi{}evi\ifmmode~\acute{c}\else \'{c}\fi{}}}]{PhysRevB.96.094510}%
  \BibitemOpen
  \bibfield  {author} {\bibinfo {author} {\bibfnamefont {J.}~\bibnamefont
  {Bekaert}}, \bibinfo {author} {\bibfnamefont {A.}~\bibnamefont {Aperis}},
  \bibinfo {author} {\bibfnamefont {B.}~\bibnamefont {Partoens}}, \bibinfo
  {author} {\bibfnamefont {P.~M.}\ \bibnamefont {Oppeneer}},\ and\ \bibinfo
  {author} {\bibfnamefont {M.~V.}\ \bibnamefont {Milo\ifmmode \check{s}\else
  \v{s}\fi{}evi\ifmmode~\acute{c}\else \'{c}\fi{}}},\ }\bibfield  {title}
  {\bibinfo {title} {{Evolution of multigap superconductivity in the atomically
  thin limit: Strain-enhanced three-gap superconductivity in monolayer
  ${\mathrm{MgB}}_{2}$}},\ }\href {https://doi.org/10.1103/PhysRevB.96.094510}
  {\bibfield  {journal} {\bibinfo  {journal} {Phys. Rev. B}\ }\textbf {\bibinfo
  {volume} {96}},\ \bibinfo {pages} {094510} (\bibinfo {year}
  {2017})}\BibitemShut {NoStop}%
\bibitem [{\citenamefont {Zhang}\ \emph {et~al.}(2016)\citenamefont {Zhang},
  \citenamefont {Gao},\ and\ \citenamefont {Dong}}]{PhysRevB.93.155430}%
  \BibitemOpen
  \bibfield  {author} {\bibinfo {author} {\bibfnamefont {J.-J.}\ \bibnamefont
  {Zhang}}, \bibinfo {author} {\bibfnamefont {B.}~\bibnamefont {Gao}},\ and\
  \bibinfo {author} {\bibfnamefont {S.}~\bibnamefont {Dong}},\ }\bibfield
  {title} {\bibinfo {title} {Strain-enhanced superconductivity of
  $\mathrm{Mo}{X}_{2}(x=\text{S} \text{or Se})$ bilayers with na
  intercalation},\ }\href {https://doi.org/10.1103/PhysRevB.93.155430}
  {\bibfield  {journal} {\bibinfo  {journal} {Phys. Rev. B}\ }\textbf {\bibinfo
  {volume} {93}},\ \bibinfo {pages} {155430} (\bibinfo {year}
  {2016})}\BibitemShut {NoStop}%
\bibitem [{\citenamefont {Lian}\ \emph {et~al.}(2017)\citenamefont {Lian},
  \citenamefont {Si}, \citenamefont {Wu},\ and\ \citenamefont
  {Duan}}]{PhysRevB.96.235426}%
  \BibitemOpen
  \bibfield  {author} {\bibinfo {author} {\bibfnamefont {C.-S.}\ \bibnamefont
  {Lian}}, \bibinfo {author} {\bibfnamefont {C.}~\bibnamefont {Si}}, \bibinfo
  {author} {\bibfnamefont {J.}~\bibnamefont {Wu}},\ and\ \bibinfo {author}
  {\bibfnamefont {W.}~\bibnamefont {Duan}},\ }\bibfield  {title} {\bibinfo
  {title} {{First-principles study of Na-intercalated bilayer
  ${\mathrm{NbSe}}_{2}$: Suppressed charge-density wave and strain-enhanced
  superconductivity}},\ }\href {https://doi.org/10.1103/PhysRevB.96.235426}
  {\bibfield  {journal} {\bibinfo  {journal} {Phys. Rev. B}\ }\textbf {\bibinfo
  {volume} {96}},\ \bibinfo {pages} {235426} (\bibinfo {year}
  {2017})}\BibitemShut {NoStop}%
\bibitem [{\citenamefont {Peng}\ \emph {et~al.}(2014)\citenamefont {Peng},
  \citenamefont {Xu}, \citenamefont {Tan}, \citenamefont {Cao}, \citenamefont
  {Xia}, \citenamefont {Shen}, \citenamefont {Huang}, \citenamefont {Wen},
  \citenamefont {Song}, \citenamefont {Zhang} \emph {et~al.}}]{peng2014tuning}%
  \BibitemOpen
  \bibfield  {author} {\bibinfo {author} {\bibfnamefont {R.}~\bibnamefont
  {Peng}}, \bibinfo {author} {\bibfnamefont {H.}~\bibnamefont {Xu}}, \bibinfo
  {author} {\bibfnamefont {S.}~\bibnamefont {Tan}}, \bibinfo {author}
  {\bibfnamefont {H.}~\bibnamefont {Cao}}, \bibinfo {author} {\bibfnamefont
  {M.}~\bibnamefont {Xia}}, \bibinfo {author} {\bibfnamefont {X.}~\bibnamefont
  {Shen}}, \bibinfo {author} {\bibfnamefont {Z.}~\bibnamefont {Huang}},
  \bibinfo {author} {\bibfnamefont {C.}~\bibnamefont {Wen}}, \bibinfo {author}
  {\bibfnamefont {Q.}~\bibnamefont {Song}}, \bibinfo {author} {\bibfnamefont
  {T.}~\bibnamefont {Zhang}}, \emph {et~al.},\ }\bibfield  {title} {\bibinfo
  {title} {{Tuning the band structure and superconductivity in single-layer
  FeSe by interface engineering}},\ }\href {https://doi.org/10.1038/ncomms6044}
  {\bibfield  {journal} {\bibinfo  {journal} {Nat. Commun.}\ }\textbf {\bibinfo
  {volume} {5}},\ \bibinfo {pages} {5044} (\bibinfo {year} {2014})}\BibitemShut
  {NoStop}%
\bibitem [{\citenamefont {Ding}\ \emph {et~al.}(2022)\citenamefont {Ding},
  \citenamefont {Qu}, \citenamefont {Han}, \citenamefont {Han}, \citenamefont
  {Zhuang}, \citenamefont {Yu}, \citenamefont {Niu}, \citenamefont {Wang},
  \citenamefont {Li}, \citenamefont {Gan}, \citenamefont {Wu},\ and\
  \citenamefont {Lu}}]{doi:10.1021/acs.nanolett.2c02947}%
  \BibitemOpen
  \bibfield  {author} {\bibinfo {author} {\bibfnamefont {D.}~\bibnamefont
  {Ding}}, \bibinfo {author} {\bibfnamefont {Z.}~\bibnamefont {Qu}}, \bibinfo
  {author} {\bibfnamefont {X.}~\bibnamefont {Han}}, \bibinfo {author}
  {\bibfnamefont {C.}~\bibnamefont {Han}}, \bibinfo {author} {\bibfnamefont
  {Q.}~\bibnamefont {Zhuang}}, \bibinfo {author} {\bibfnamefont {X.-L.}\
  \bibnamefont {Yu}}, \bibinfo {author} {\bibfnamefont {R.}~\bibnamefont
  {Niu}}, \bibinfo {author} {\bibfnamefont {Z.}~\bibnamefont {Wang}}, \bibinfo
  {author} {\bibfnamefont {Z.}~\bibnamefont {Li}}, \bibinfo {author}
  {\bibfnamefont {Z.}~\bibnamefont {Gan}}, \bibinfo {author} {\bibfnamefont
  {J.}~\bibnamefont {Wu}},\ and\ \bibinfo {author} {\bibfnamefont
  {J.}~\bibnamefont {Lu}},\ }\bibfield  {title} {\bibinfo {title} {Multivalley
  superconductivity in monolayer transition metal dichalcogenides},\ }\href
  {https://doi.org/10.1021/acs.nanolett.2c02947} {\bibfield  {journal}
  {\bibinfo  {journal} {Nano Lett.}\ }\textbf {\bibinfo {volume} {22}},\
  \bibinfo {pages} {7919} (\bibinfo {year} {2022})}\BibitemShut {NoStop}%
\bibitem [{\citenamefont {Telford}\ \emph {et~al.}(2020)\citenamefont
  {Telford}, \citenamefont {Russell}, \citenamefont {Swann}, \citenamefont
  {Fowler}, \citenamefont {Wang}, \citenamefont {Lee}, \citenamefont
  {Zangiabadi}, \citenamefont {Watanabe}, \citenamefont {Taniguchi},
  \citenamefont {Nuckolls}, \citenamefont {Batail}, \citenamefont {Zhu},
  \citenamefont {Malen}, \citenamefont {Dean},\ and\ \citenamefont
  {Roy}}]{doi:10.1021/acs.nanolett.9b04891}%
  \BibitemOpen
  \bibfield  {author} {\bibinfo {author} {\bibfnamefont {E.~J.}\ \bibnamefont
  {Telford}}, \bibinfo {author} {\bibfnamefont {J.~C.}\ \bibnamefont
  {Russell}}, \bibinfo {author} {\bibfnamefont {J.~R.}\ \bibnamefont {Swann}},
  \bibinfo {author} {\bibfnamefont {B.}~\bibnamefont {Fowler}}, \bibinfo
  {author} {\bibfnamefont {X.}~\bibnamefont {Wang}}, \bibinfo {author}
  {\bibfnamefont {K.}~\bibnamefont {Lee}}, \bibinfo {author} {\bibfnamefont
  {A.}~\bibnamefont {Zangiabadi}}, \bibinfo {author} {\bibfnamefont
  {K.}~\bibnamefont {Watanabe}}, \bibinfo {author} {\bibfnamefont
  {T.}~\bibnamefont {Taniguchi}}, \bibinfo {author} {\bibfnamefont
  {C.}~\bibnamefont {Nuckolls}}, \bibinfo {author} {\bibfnamefont
  {P.}~\bibnamefont {Batail}}, \bibinfo {author} {\bibfnamefont
  {X.}~\bibnamefont {Zhu}}, \bibinfo {author} {\bibfnamefont {J.~A.}\
  \bibnamefont {Malen}}, \bibinfo {author} {\bibfnamefont {C.~R.}\ \bibnamefont
  {Dean}},\ and\ \bibinfo {author} {\bibfnamefont {X.}~\bibnamefont {Roy}},\
  }\bibfield  {title} {\bibinfo {title} {{Doping-Induced Superconductivity in
  the van der Waals Superatomic Crystal Re$_6$Se$_8$Cl$_2$}},\ }\href
  {https://doi.org/10.1021/acs.nanolett.9b04891} {\bibfield  {journal}
  {\bibinfo  {journal} {Nano Lett.}\ }\textbf {\bibinfo {volume} {20}},\
  \bibinfo {pages} {1718} (\bibinfo {year} {2020})}\BibitemShut {NoStop}%
\bibitem [{\citenamefont {Li}\ \emph {et~al.}(2019)\citenamefont {Li},
  \citenamefont {Stavrou}, \citenamefont {Zhu}, \citenamefont {Clarke},
  \citenamefont {Li},\ and\ \citenamefont {Huang}}]{PhysRevB.99.220503}%
  \BibitemOpen
  \bibfield  {author} {\bibinfo {author} {\bibfnamefont {Y.-L.}\ \bibnamefont
  {Li}}, \bibinfo {author} {\bibfnamefont {E.}~\bibnamefont {Stavrou}},
  \bibinfo {author} {\bibfnamefont {Q.}~\bibnamefont {Zhu}}, \bibinfo {author}
  {\bibfnamefont {S.~M.}\ \bibnamefont {Clarke}}, \bibinfo {author}
  {\bibfnamefont {Y.}~\bibnamefont {Li}},\ and\ \bibinfo {author}
  {\bibfnamefont {H.-M.}\ \bibnamefont {Huang}},\ }\bibfield  {title} {\bibinfo
  {title} {{Superconductivity in the van der Waals layered compound
  ${\mathrm{PS}}_{2}$}},\ }\href {https://doi.org/10.1103/PhysRevB.99.220503}
  {\bibfield  {journal} {\bibinfo  {journal} {Phys. Rev. B}\ }\textbf {\bibinfo
  {volume} {99}},\ \bibinfo {pages} {220503} (\bibinfo {year}
  {2019})}\BibitemShut {NoStop}%
\bibitem [{\citenamefont {Baidya}\ \emph {et~al.}(2021)\citenamefont {Baidya},
  \citenamefont {Sahani}, \citenamefont {Kundu}, \citenamefont {Kaur},
  \citenamefont {Tiwari}, \citenamefont {Bagwe}, \citenamefont {Jesudasan},
  \citenamefont {Narayan}, \citenamefont {Raychaudhuri},\ and\ \citenamefont
  {Bid}}]{PhysRevB.104.174510}%
  \BibitemOpen
  \bibfield  {author} {\bibinfo {author} {\bibfnamefont {P.}~\bibnamefont
  {Baidya}}, \bibinfo {author} {\bibfnamefont {D.}~\bibnamefont {Sahani}},
  \bibinfo {author} {\bibfnamefont {H.~K.}\ \bibnamefont {Kundu}}, \bibinfo
  {author} {\bibfnamefont {S.}~\bibnamefont {Kaur}}, \bibinfo {author}
  {\bibfnamefont {P.}~\bibnamefont {Tiwari}}, \bibinfo {author} {\bibfnamefont
  {V.}~\bibnamefont {Bagwe}}, \bibinfo {author} {\bibfnamefont
  {J.}~\bibnamefont {Jesudasan}}, \bibinfo {author} {\bibfnamefont
  {A.}~\bibnamefont {Narayan}}, \bibinfo {author} {\bibfnamefont
  {P.}~\bibnamefont {Raychaudhuri}},\ and\ \bibinfo {author} {\bibfnamefont
  {A.}~\bibnamefont {Bid}},\ }\bibfield  {title} {\bibinfo {title} {{Transition
  from three- to two-dimensional Ising superconductivity in few-layer
  ${\mathrm{NbSe}}_{2}$ by proximity effect from van der Waals
  heterostacking}},\ }\href {https://doi.org/10.1103/PhysRevB.104.174510}
  {\bibfield  {journal} {\bibinfo  {journal} {Phys. Rev. B}\ }\textbf {\bibinfo
  {volume} {104}},\ \bibinfo {pages} {174510} (\bibinfo {year}
  {2021})}\BibitemShut {NoStop}%
\bibitem [{\citenamefont {Navarro-Moratalla}\ \emph {et~al.}(2016)\citenamefont
  {Navarro-Moratalla}, \citenamefont {Island}, \citenamefont {Manas-Valero},
  \citenamefont {Pinilla-Cienfuegos}, \citenamefont {Castellanos-Gomez},
  \citenamefont {Quereda}, \citenamefont {Rubio-Bollinger}, \citenamefont
  {Chirolli}, \citenamefont {Silva-Guill{\'e}n}, \citenamefont {Agra{\"\i}t}
  \emph {et~al.}}]{navarro2016enhanced}%
  \BibitemOpen
  \bibfield  {author} {\bibinfo {author} {\bibfnamefont {E.}~\bibnamefont
  {Navarro-Moratalla}}, \bibinfo {author} {\bibfnamefont {J.~O.}\ \bibnamefont
  {Island}}, \bibinfo {author} {\bibfnamefont {S.}~\bibnamefont
  {Manas-Valero}}, \bibinfo {author} {\bibfnamefont {E.}~\bibnamefont
  {Pinilla-Cienfuegos}}, \bibinfo {author} {\bibfnamefont {A.}~\bibnamefont
  {Castellanos-Gomez}}, \bibinfo {author} {\bibfnamefont {J.}~\bibnamefont
  {Quereda}}, \bibinfo {author} {\bibfnamefont {G.}~\bibnamefont
  {Rubio-Bollinger}}, \bibinfo {author} {\bibfnamefont {L.}~\bibnamefont
  {Chirolli}}, \bibinfo {author} {\bibfnamefont {J.~A.}\ \bibnamefont
  {Silva-Guill{\'e}n}}, \bibinfo {author} {\bibfnamefont {N.}~\bibnamefont
  {Agra{\"\i}t}}, \emph {et~al.},\ }\bibfield  {title} {\bibinfo {title}
  {{Enhanced superconductivity in atomically thin TaS$_2$}},\ }\href
  {https://doi.org/10.1038/ncomms11043} {\bibfield  {journal} {\bibinfo
  {journal} {Nat. Commun.}\ }\textbf {\bibinfo {volume} {7}},\ \bibinfo {pages}
  {11043} (\bibinfo {year} {2016})}\BibitemShut {NoStop}%
\bibitem [{\citenamefont {De~la Barrera}\ \emph {et~al.}(2018)\citenamefont
  {De~la Barrera}, \citenamefont {Sinko}, \citenamefont {Gopalan},
  \citenamefont {Sivadas}, \citenamefont {Seyler}, \citenamefont {Watanabe},
  \citenamefont {Taniguchi}, \citenamefont {Tsen}, \citenamefont {Xu},
  \citenamefont {Xiao} \emph {et~al.}}]{de2018tuning}%
  \BibitemOpen
  \bibfield  {author} {\bibinfo {author} {\bibfnamefont {S.~C.}\ \bibnamefont
  {De~la Barrera}}, \bibinfo {author} {\bibfnamefont {M.~R.}\ \bibnamefont
  {Sinko}}, \bibinfo {author} {\bibfnamefont {D.~P.}\ \bibnamefont {Gopalan}},
  \bibinfo {author} {\bibfnamefont {N.}~\bibnamefont {Sivadas}}, \bibinfo
  {author} {\bibfnamefont {K.~L.}\ \bibnamefont {Seyler}}, \bibinfo {author}
  {\bibfnamefont {K.}~\bibnamefont {Watanabe}}, \bibinfo {author}
  {\bibfnamefont {T.}~\bibnamefont {Taniguchi}}, \bibinfo {author}
  {\bibfnamefont {A.~W.}\ \bibnamefont {Tsen}}, \bibinfo {author}
  {\bibfnamefont {X.}~\bibnamefont {Xu}}, \bibinfo {author} {\bibfnamefont
  {D.}~\bibnamefont {Xiao}}, \emph {et~al.},\ }\bibfield  {title} {\bibinfo
  {title} {{Tuning Ising superconductivity with layer and spin--orbit coupling
  in two-dimensional transition-metal dichalcogenides}},\ }\href
  {https://doi.org/10.1038/s41467-018-03888-4} {\bibfield  {journal} {\bibinfo
  {journal} {Nat. Commun.}\ }\textbf {\bibinfo {volume} {9}},\ \bibinfo {pages}
  {1427} (\bibinfo {year} {2018})}\BibitemShut {NoStop}%
\bibitem [{\citenamefont {Lian}\ \emph {et~al.}(2018)\citenamefont {Lian},
  \citenamefont {Si},\ and\ \citenamefont
  {Duan}}]{doi:10.1021/acs.nanolett.8b00237}%
  \BibitemOpen
  \bibfield  {author} {\bibinfo {author} {\bibfnamefont {C.-S.}\ \bibnamefont
  {Lian}}, \bibinfo {author} {\bibfnamefont {C.}~\bibnamefont {Si}},\ and\
  \bibinfo {author} {\bibfnamefont {W.}~\bibnamefont {Duan}},\ }\bibfield
  {title} {\bibinfo {title} {{Unveiling Charge-Density Wave, Superconductivity,
  and Their Competitive Nature in Two-Dimensional NbSe$_2$}},\ }\href
  {https://doi.org/10.1021/acs.nanolett.8b00237} {\bibfield  {journal}
  {\bibinfo  {journal} {Nano Lett.}\ }\textbf {\bibinfo {volume} {18}},\
  \bibinfo {pages} {2924} (\bibinfo {year} {2018})}\BibitemShut {NoStop}%
\bibitem [{\citenamefont {Qiu}\ \emph {et~al.}(2022)\citenamefont {Qiu},
  \citenamefont {Zhang}, \citenamefont {Yang}, \citenamefont {Lu},\ and\
  \citenamefont {Liu}}]{PhysRevB.105.165101}%
  \BibitemOpen
  \bibfield  {author} {\bibinfo {author} {\bibfnamefont {X.-L.}\ \bibnamefont
  {Qiu}}, \bibinfo {author} {\bibfnamefont {J.-F.}\ \bibnamefont {Zhang}},
  \bibinfo {author} {\bibfnamefont {H.-C.}\ \bibnamefont {Yang}}, \bibinfo
  {author} {\bibfnamefont {Z.-Y.}\ \bibnamefont {Lu}},\ and\ \bibinfo {author}
  {\bibfnamefont {K.}~\bibnamefont {Liu}},\ }\bibfield  {title} {\bibinfo
  {title} {{Superconductivity in monolayer ${\mathrm{Ba}}_{2}\mathrm{N}$
  electride: First-principles study}},\ }\href
  {https://doi.org/10.1103/PhysRevB.105.165101} {\bibfield  {journal} {\bibinfo
   {journal} {Phys. Rev. B}\ }\textbf {\bibinfo {volume} {105}},\ \bibinfo
  {pages} {165101} (\bibinfo {year} {2022})}\BibitemShut {NoStop}%
\bibitem [{\citenamefont {Campi}\ \emph {et~al.}(2021)\citenamefont {Campi},
  \citenamefont {Kumari},\ and\ \citenamefont {Marzari}}]{campi2021prediction}%
  \BibitemOpen
  \bibfield  {author} {\bibinfo {author} {\bibfnamefont {D.}~\bibnamefont
  {Campi}}, \bibinfo {author} {\bibfnamefont {S.}~\bibnamefont {Kumari}},\ and\
  \bibinfo {author} {\bibfnamefont {N.}~\bibnamefont {Marzari}},\ }\bibfield
  {title} {\bibinfo {title} {{Prediction of phonon-mediated superconductivity
  with high critical temperature in the two-dimensional topological semimetal
  W$_2$N$_3$}},\ }\href
  {https://doi.org/https://doi.org/10.1021/acs.nanolett.0c05125} {\bibfield
  {journal} {\bibinfo  {journal} {Nano Lett.}\ }\textbf {\bibinfo {volume}
  {21}},\ \bibinfo {pages} {3435} (\bibinfo {year} {2021})}\BibitemShut
  {NoStop}%
\bibitem [{\citenamefont {Chen}\ and\ \citenamefont
  {Ge}(2021)}]{PhysRevB.103.064510}%
  \BibitemOpen
  \bibfield  {author} {\bibinfo {author} {\bibfnamefont {J.}~\bibnamefont
  {Chen}}\ and\ \bibinfo {author} {\bibfnamefont {Y.}~\bibnamefont {Ge}},\
  }\bibfield  {title} {\bibinfo {title} {{Emergence of intrinsic
  superconductivity in monolayer ${\mathrm{W}}_{2}{\mathrm{N}}_{3}$}},\ }\href
  {https://doi.org/10.1103/PhysRevB.103.064510} {\bibfield  {journal} {\bibinfo
   {journal} {Phys. Rev. B}\ }\textbf {\bibinfo {volume} {103}},\ \bibinfo
  {pages} {064510} (\bibinfo {year} {2021})}\BibitemShut {NoStop}%
\bibitem [{\citenamefont {Guo}\ \emph {et~al.}(2019)\citenamefont {Guo},
  \citenamefont {Peng}, \citenamefont {Qin}, \citenamefont {Zeng},
  \citenamefont {Zhao}, \citenamefont {Wu}, \citenamefont {Chu}, \citenamefont
  {Wang}, \citenamefont {Wu},\ and\ \citenamefont {Xie}}]{guo2019freestanding}%
  \BibitemOpen
  \bibfield  {author} {\bibinfo {author} {\bibfnamefont {Y.}~\bibnamefont
  {Guo}}, \bibinfo {author} {\bibfnamefont {J.}~\bibnamefont {Peng}}, \bibinfo
  {author} {\bibfnamefont {W.}~\bibnamefont {Qin}}, \bibinfo {author}
  {\bibfnamefont {J.}~\bibnamefont {Zeng}}, \bibinfo {author} {\bibfnamefont
  {J.}~\bibnamefont {Zhao}}, \bibinfo {author} {\bibfnamefont {J.}~\bibnamefont
  {Wu}}, \bibinfo {author} {\bibfnamefont {W.}~\bibnamefont {Chu}}, \bibinfo
  {author} {\bibfnamefont {L.}~\bibnamefont {Wang}}, \bibinfo {author}
  {\bibfnamefont {C.}~\bibnamefont {Wu}},\ and\ \bibinfo {author}
  {\bibfnamefont {Y.}~\bibnamefont {Xie}},\ }\bibfield  {title} {\bibinfo
  {title} {{Freestanding cubic ZrN single-crystalline films with
  two-dimensional superconductivity}},\ }\href
  {https://doi.org/https://doi.org/10.1021/jacs.9b05114} {\bibfield  {journal}
  {\bibinfo  {journal} {J. Am. Chem. Soc.}\ }\textbf {\bibinfo {volume}
  {141}},\ \bibinfo {pages} {10183} (\bibinfo {year} {2019})}\BibitemShut
  {NoStop}%
\bibitem [{\citenamefont {Pereira}\ \emph {et~al.}(2022)\citenamefont
  {Pereira}, \citenamefont {Faccin},\ and\ \citenamefont
  {da~Silva}}]{D2NR00395C}%
  \BibitemOpen
  \bibfield  {author} {\bibinfo {author} {\bibfnamefont {Z.~S.}\ \bibnamefont
  {Pereira}}, \bibinfo {author} {\bibfnamefont {G.~M.}\ \bibnamefont
  {Faccin}},\ and\ \bibinfo {author} {\bibfnamefont {E.~Z.}\ \bibnamefont
  {da~Silva}},\ }\bibfield  {title} {\bibinfo {title} {{Strain-induced multigap
  superconductivity in electrene Mo$_2$N: a first principles study}},\ }\href
  {https://doi.org/10.1039/D2NR00395C} {\bibfield  {journal} {\bibinfo
  {journal} {Nanoscale}\ }\textbf {\bibinfo {volume} {14}},\ \bibinfo {pages}
  {8594} (\bibinfo {year} {2022})}\BibitemShut {NoStop}%
\bibitem [{\citenamefont {Yan}\ \emph {et~al.}(2021)\citenamefont {Yan},
  \citenamefont {Wang}, \citenamefont {Huang}, \citenamefont {Li},
  \citenamefont {Xue}, \citenamefont {Zhang}, \citenamefont {Ren},\ and\
  \citenamefont {Zhou}}]{yan2021surface}%
  \BibitemOpen
  \bibfield  {author} {\bibinfo {author} {\bibfnamefont {L.}~\bibnamefont
  {Yan}}, \bibinfo {author} {\bibfnamefont {B.-T.}\ \bibnamefont {Wang}},
  \bibinfo {author} {\bibfnamefont {X.}~\bibnamefont {Huang}}, \bibinfo
  {author} {\bibfnamefont {Q.}~\bibnamefont {Li}}, \bibinfo {author}
  {\bibfnamefont {K.}~\bibnamefont {Xue}}, \bibinfo {author} {\bibfnamefont
  {J.}~\bibnamefont {Zhang}}, \bibinfo {author} {\bibfnamefont
  {W.}~\bibnamefont {Ren}},\ and\ \bibinfo {author} {\bibfnamefont
  {L.}~\bibnamefont {Zhou}},\ }\bibfield  {title} {\bibinfo {title} {{Surface
  passivation induced a significant enhancement of superconductivity in layered
  two-dimensional MSi$_2$N$_4$ (M= Ta and Nb) materials}},\ }\href
  {https://doi.org/https://doi.org/10.1039/D1NR05560G} {\bibfield  {journal}
  {\bibinfo  {journal} {Nanoscale}\ }\textbf {\bibinfo {volume} {13}},\
  \bibinfo {pages} {18947} (\bibinfo {year} {2021})}\BibitemShut {NoStop}%
\bibitem [{\citenamefont {Cooper}\ \emph {et~al.}(1970)\citenamefont {Cooper},
  \citenamefont {Corenzwit}, \citenamefont {Longinotti}, \citenamefont
  {Matthias},\ and\ \citenamefont {Zachariasen}}]{doi:10.1073/pnas.67.1.313}%
  \BibitemOpen
  \bibfield  {author} {\bibinfo {author} {\bibfnamefont {A.~S.}\ \bibnamefont
  {Cooper}}, \bibinfo {author} {\bibfnamefont {E.}~\bibnamefont {Corenzwit}},
  \bibinfo {author} {\bibfnamefont {L.~D.}\ \bibnamefont {Longinotti}},
  \bibinfo {author} {\bibfnamefont {B.~T.}\ \bibnamefont {Matthias}},\ and\
  \bibinfo {author} {\bibfnamefont {W.~H.}\ \bibnamefont {Zachariasen}},\
  }\bibfield  {title} {\bibinfo {title} {{Superconductivity: The Transition
  Temperature Peak Below Four Electrons per Atom}},\ }\href
  {https://doi.org/10.1073/pnas.67.1.313} {\bibfield  {journal} {\bibinfo
  {journal} {Proc. Natl. Acad. Sci.}\ }\textbf {\bibinfo {volume} {67}},\
  \bibinfo {pages} {313} (\bibinfo {year} {1970})}\BibitemShut {NoStop}%
\bibitem [{\citenamefont {Puthirath~Balan}\ \emph {et~al.}(2018)\citenamefont
  {Puthirath~Balan}, \citenamefont {Radhakrishnan}, \citenamefont {Woellner},
  \citenamefont {Sinha}, \citenamefont {Deng}, \citenamefont {Reyes},
  \citenamefont {Rao}, \citenamefont {Paulose}, \citenamefont {Neupane},
  \citenamefont {Apte} \emph {et~al.}}]{puthirath2018exfoliation}%
  \BibitemOpen
  \bibfield  {author} {\bibinfo {author} {\bibfnamefont {A.}~\bibnamefont
  {Puthirath~Balan}}, \bibinfo {author} {\bibfnamefont {S.}~\bibnamefont
  {Radhakrishnan}}, \bibinfo {author} {\bibfnamefont {C.~F.}\ \bibnamefont
  {Woellner}}, \bibinfo {author} {\bibfnamefont {S.~K.}\ \bibnamefont {Sinha}},
  \bibinfo {author} {\bibfnamefont {L.}~\bibnamefont {Deng}}, \bibinfo {author}
  {\bibfnamefont {C.~d.~l.}\ \bibnamefont {Reyes}}, \bibinfo {author}
  {\bibfnamefont {B.~M.}\ \bibnamefont {Rao}}, \bibinfo {author} {\bibfnamefont
  {M.}~\bibnamefont {Paulose}}, \bibinfo {author} {\bibfnamefont
  {R.}~\bibnamefont {Neupane}}, \bibinfo {author} {\bibfnamefont
  {A.}~\bibnamefont {Apte}}, \emph {et~al.},\ }\bibfield  {title} {\bibinfo
  {title} {{Exfoliation of a non-van der Waals material from iron ore
  hematite}},\ }\href
  {https://doi.org/https://doi.org/10.1038/s41565-018-0134-y} {\bibfield
  {journal} {\bibinfo  {journal} {Nat. Nanotech.}\ }\textbf {\bibinfo {volume}
  {13}},\ \bibinfo {pages} {602} (\bibinfo {year} {2018})}\BibitemShut
  {NoStop}%
\bibitem [{\citenamefont {Jiang}\ \emph
  {et~al.}(2023{\natexlab{a}})\citenamefont {Jiang}, \citenamefont {Ji},
  \citenamefont {Gong}, \citenamefont {Ding}, \citenamefont {Li}, \citenamefont
  {Li}, \citenamefont {Li},\ and\ \citenamefont {Geng}}]{jiang2023mechanical}%
  \BibitemOpen
  \bibfield  {author} {\bibinfo {author} {\bibfnamefont {K.}~\bibnamefont
  {Jiang}}, \bibinfo {author} {\bibfnamefont {J.}~\bibnamefont {Ji}}, \bibinfo
  {author} {\bibfnamefont {W.}~\bibnamefont {Gong}}, \bibinfo {author}
  {\bibfnamefont {L.}~\bibnamefont {Ding}}, \bibinfo {author} {\bibfnamefont
  {J.}~\bibnamefont {Li}}, \bibinfo {author} {\bibfnamefont {P.}~\bibnamefont
  {Li}}, \bibinfo {author} {\bibfnamefont {B.}~\bibnamefont {Li}},\ and\
  \bibinfo {author} {\bibfnamefont {F.}~\bibnamefont {Geng}},\ }\bibfield
  {title} {\bibinfo {title} {{Mechanical cleavage of non-van der Waals
  structures towards two-dimensional crystals}},\ }\href
  {https://doi.org/https://doi.org/10.1038/s44160-022-00182-6} {\bibfield
  {journal} {\bibinfo  {journal} {Nat. Synth}\ }\textbf {\bibinfo {volume}
  {2}},\ \bibinfo {pages} {58} (\bibinfo {year}
  {2023}{\natexlab{a}})}\BibitemShut {NoStop}%
\bibitem [{\citenamefont {Niu}\ \emph {et~al.}(2023)\citenamefont {Niu},
  \citenamefont {Zhang}, \citenamefont {Cui}, \citenamefont {Wu},\ and\
  \citenamefont {Yang}}]{doi:10.1021/acs.nanolett.3c00113}%
  \BibitemOpen
  \bibfield  {author} {\bibinfo {author} {\bibfnamefont {Y.}~\bibnamefont
  {Niu}}, \bibinfo {author} {\bibfnamefont {K.}~\bibnamefont {Zhang}}, \bibinfo
  {author} {\bibfnamefont {X.}~\bibnamefont {Cui}}, \bibinfo {author}
  {\bibfnamefont {X.}~\bibnamefont {Wu}},\ and\ \bibinfo {author}
  {\bibfnamefont {J.}~\bibnamefont {Yang}},\ }\bibfield  {title} {\bibinfo
  {title} {{Two-Dimensional Iron Silicide (FeSi$_x$) Alloys with
  Above-Room-Temperature Ferromagnetism}},\ }\href
  {https://doi.org/10.1021/acs.nanolett.3c00113} {\bibfield  {journal}
  {\bibinfo  {journal} {Nano Lett.}\ }\textbf {\bibinfo {volume} {23}},\
  \bibinfo {pages} {2332} (\bibinfo {year} {2023})}\BibitemShut {NoStop}%
\bibitem [{\citenamefont {Friedrich}\ \emph {et~al.}(2022)\citenamefont
  {Friedrich}, \citenamefont {Ghorbani-Asl}, \citenamefont {Curtarolo},\ and\
  \citenamefont {Krasheninnikov}}]{doi:10.1021/acs.nanolett.1c03841}%
  \BibitemOpen
  \bibfield  {author} {\bibinfo {author} {\bibfnamefont {R.}~\bibnamefont
  {Friedrich}}, \bibinfo {author} {\bibfnamefont {M.}~\bibnamefont
  {Ghorbani-Asl}}, \bibinfo {author} {\bibfnamefont {S.}~\bibnamefont
  {Curtarolo}},\ and\ \bibinfo {author} {\bibfnamefont {A.~V.}\ \bibnamefont
  {Krasheninnikov}},\ }\bibfield  {title} {\bibinfo {title} {{Data-Driven Quest
  for Two-Dimensional Non-van der Waals Materials}},\ }\href
  {https://doi.org/10.1021/acs.nanolett.1c03841} {\bibfield  {journal}
  {\bibinfo  {journal} {Nano Lett.}\ }\textbf {\bibinfo {volume} {22}},\
  \bibinfo {pages} {989} (\bibinfo {year} {2022})}\BibitemShut {NoStop}%
\bibitem [{\citenamefont {Gao}\ \emph {et~al.}(2025)\citenamefont {Gao},
  \citenamefont {Zhou}, \citenamefont {Ping}, \citenamefont {Wang},
  \citenamefont {Hung}, \citenamefont {Cao}, \citenamefont {Geiwitz},
  \citenamefont {Natale}, \citenamefont {Lin}, \citenamefont {Burch},
  \citenamefont {Saito}, \citenamefont {Terrones},\ and\ \citenamefont
  {Ling}}]{doi:10.1021/acsnano.4c12155}%
  \BibitemOpen
  \bibfield  {author} {\bibinfo {author} {\bibfnamefont {H.}~\bibnamefont
  {Gao}}, \bibinfo {author} {\bibfnamefont {D.}~\bibnamefont {Zhou}}, \bibinfo
  {author} {\bibfnamefont {L.}~\bibnamefont {Ping}}, \bibinfo {author}
  {\bibfnamefont {Z.}~\bibnamefont {Wang}}, \bibinfo {author} {\bibfnamefont
  {N.~T.}\ \bibnamefont {Hung}}, \bibinfo {author} {\bibfnamefont
  {J.}~\bibnamefont {Cao}}, \bibinfo {author} {\bibfnamefont {M.}~\bibnamefont
  {Geiwitz}}, \bibinfo {author} {\bibfnamefont {G.}~\bibnamefont {Natale}},
  \bibinfo {author} {\bibfnamefont {Y.~C.}\ \bibnamefont {Lin}}, \bibinfo
  {author} {\bibfnamefont {K.~S.}\ \bibnamefont {Burch}}, \bibinfo {author}
  {\bibfnamefont {R.}~\bibnamefont {Saito}}, \bibinfo {author} {\bibfnamefont
  {M.}~\bibnamefont {Terrones}},\ and\ \bibinfo {author} {\bibfnamefont
  {X.}~\bibnamefont {Ling}},\ }\bibfield  {title} {\bibinfo {title}
  {{Downscaling of Non-Van der Waals Semimetallic W$_5$N$_6$ with Resistivity
  Preservation}},\ }\href {https://doi.org/10.1021/acsnano.4c12155} {\bibfield
  {journal} {\bibinfo  {journal} {ACS Nano}\ }\textbf {\bibinfo {volume}
  {19}},\ \bibinfo {pages} {3362} (\bibinfo {year} {2025})}\BibitemShut
  {NoStop}%
\bibitem [{\citenamefont {Kresse}\ and\ \citenamefont
  {Furthm{\"u}ller}(1996{\natexlab{a}})}]{kresse1996efficiency}%
  \BibitemOpen
  \bibfield  {author} {\bibinfo {author} {\bibfnamefont {G.}~\bibnamefont
  {Kresse}}\ and\ \bibinfo {author} {\bibfnamefont {J.}~\bibnamefont
  {Furthm{\"u}ller}},\ }\bibfield  {title} {\bibinfo {title} {Efficiency of
  ab-initio total energy calculations for metals and semiconductors using a
  plane-wave basis set},\ }\href {https://doi.org/10.1016/0927-0256(96)00008-0}
  {\bibfield  {journal} {\bibinfo  {journal} {Comput. Mater. Sci}\ }\textbf
  {\bibinfo {volume} {6}},\ \bibinfo {pages} {15} (\bibinfo {year}
  {1996}{\natexlab{a}})}\BibitemShut {NoStop}%
\bibitem [{\citenamefont {Kresse}\ and\ \citenamefont
  {Furthm{\"u}ller}(1996{\natexlab{b}})}]{kresse1996efficient}%
  \BibitemOpen
  \bibfield  {author} {\bibinfo {author} {\bibfnamefont {G.}~\bibnamefont
  {Kresse}}\ and\ \bibinfo {author} {\bibfnamefont {J.}~\bibnamefont
  {Furthm{\"u}ller}},\ }\bibfield  {title} {\bibinfo {title} {Efficient
  iterative schemes for ab initio total-energy calculations using a plane-wave
  basis set},\ }\href
  {https://doi.org/https://doi.org/10.1103/PhysRevB.54.11169} {\bibfield
  {journal} {\bibinfo  {journal} {Phys. Rev. B}\ }\textbf {\bibinfo {volume}
  {54}},\ \bibinfo {pages} {11169} (\bibinfo {year}
  {1996}{\natexlab{b}})}\BibitemShut {NoStop}%
\bibitem [{\citenamefont {Kresse}\ and\ \citenamefont
  {Joubert}(1999)}]{kresse1999ultrasoft}%
  \BibitemOpen
  \bibfield  {author} {\bibinfo {author} {\bibfnamefont {G.}~\bibnamefont
  {Kresse}}\ and\ \bibinfo {author} {\bibfnamefont {D.}~\bibnamefont
  {Joubert}},\ }\bibfield  {title} {\bibinfo {title} {From ultrasoft
  pseudopotentials to the projector augmented-wave method},\ }\href
  {https://doi.org/10.1103/PhysRevB.59.1758} {\bibfield  {journal} {\bibinfo
  {journal} {Phys. Rev. B}\ }\textbf {\bibinfo {volume} {59}},\ \bibinfo
  {pages} {1758} (\bibinfo {year} {1999})}\BibitemShut {NoStop}%
\bibitem [{\citenamefont {Hamann}(2013)}]{PhysRevB.88.085117}%
  \BibitemOpen
  \bibfield  {author} {\bibinfo {author} {\bibfnamefont {D.~R.}\ \bibnamefont
  {Hamann}},\ }\bibfield  {title} {\bibinfo {title} {{Optimized norm-conserving
  Vanderbilt pseudopotentials}},\ }\href
  {https://doi.org/10.1103/PhysRevB.88.085117} {\bibfield  {journal} {\bibinfo
  {journal} {Phys. Rev. B}\ }\textbf {\bibinfo {volume} {88}},\ \bibinfo
  {pages} {085117} (\bibinfo {year} {2013})}\BibitemShut {NoStop}%
\bibitem [{\citenamefont {Giannozzi}\ \emph {et~al.}(2017)\citenamefont
  {Giannozzi}, \citenamefont {Andreussi}, \citenamefont {Brumme}, \citenamefont
  {Bunau}, \citenamefont {Nardelli}, \citenamefont {Calandra}, \citenamefont
  {Car}, \citenamefont {Cavazzoni}, \citenamefont {Ceresoli}, \citenamefont
  {Cococcioni} \emph {et~al.}}]{giannozzi2017advanced}%
  \BibitemOpen
  \bibfield  {author} {\bibinfo {author} {\bibfnamefont {P.}~\bibnamefont
  {Giannozzi}}, \bibinfo {author} {\bibfnamefont {O.}~\bibnamefont
  {Andreussi}}, \bibinfo {author} {\bibfnamefont {T.}~\bibnamefont {Brumme}},
  \bibinfo {author} {\bibfnamefont {O.}~\bibnamefont {Bunau}}, \bibinfo
  {author} {\bibfnamefont {M.~B.}\ \bibnamefont {Nardelli}}, \bibinfo {author}
  {\bibfnamefont {M.}~\bibnamefont {Calandra}}, \bibinfo {author}
  {\bibfnamefont {R.}~\bibnamefont {Car}}, \bibinfo {author} {\bibfnamefont
  {C.}~\bibnamefont {Cavazzoni}}, \bibinfo {author} {\bibfnamefont
  {D.}~\bibnamefont {Ceresoli}}, \bibinfo {author} {\bibfnamefont
  {M.}~\bibnamefont {Cococcioni}}, \emph {et~al.},\ }\bibfield  {title}
  {\bibinfo {title} {Advanced capabilities for materials modelling with quantum
  espresso},\ }\href {https://doi.org/10.1088/1361-648X/aa8f79} {\bibfield
  {journal} {\bibinfo  {journal} {J. Phys.: Conden. Matter}\ }\textbf {\bibinfo
  {volume} {29}},\ \bibinfo {pages} {465901} (\bibinfo {year}
  {2017})}\BibitemShut {NoStop}%
\bibitem [{\citenamefont {Giannozzi}\ \emph {et~al.}(2020)\citenamefont
  {Giannozzi}, \citenamefont {Baseggio}, \citenamefont {Bonfà}, \citenamefont
  {Brunato}, \citenamefont {Car}, \citenamefont {Carnimeo}, \citenamefont
  {Cavazzoni}, \citenamefont {de~Gironcoli}, \citenamefont {Delugas},
  \citenamefont {Ferrari~Ruffino}, \citenamefont {Ferretti}, \citenamefont
  {Marzari}, \citenamefont {Timrov}, \citenamefont {Urru},\ and\ \citenamefont
  {Baroni}}]{giannozzi2020quantum}%
  \BibitemOpen
  \bibfield  {author} {\bibinfo {author} {\bibfnamefont {P.}~\bibnamefont
  {Giannozzi}}, \bibinfo {author} {\bibfnamefont {O.}~\bibnamefont {Baseggio}},
  \bibinfo {author} {\bibfnamefont {P.}~\bibnamefont {Bonfà}}, \bibinfo
  {author} {\bibfnamefont {D.}~\bibnamefont {Brunato}}, \bibinfo {author}
  {\bibfnamefont {R.}~\bibnamefont {Car}}, \bibinfo {author} {\bibfnamefont
  {I.}~\bibnamefont {Carnimeo}}, \bibinfo {author} {\bibfnamefont
  {C.}~\bibnamefont {Cavazzoni}}, \bibinfo {author} {\bibfnamefont
  {S.}~\bibnamefont {de~Gironcoli}}, \bibinfo {author} {\bibfnamefont
  {P.}~\bibnamefont {Delugas}}, \bibinfo {author} {\bibfnamefont
  {F.}~\bibnamefont {Ferrari~Ruffino}}, \bibinfo {author} {\bibfnamefont
  {A.}~\bibnamefont {Ferretti}}, \bibinfo {author} {\bibfnamefont
  {N.}~\bibnamefont {Marzari}}, \bibinfo {author} {\bibfnamefont
  {I.}~\bibnamefont {Timrov}}, \bibinfo {author} {\bibfnamefont
  {A.}~\bibnamefont {Urru}},\ and\ \bibinfo {author} {\bibfnamefont
  {S.}~\bibnamefont {Baroni}},\ }\bibfield  {title} {\bibinfo {title} {{Quantum
  ESPRESSO toward the exascale}},\ }\href {https://doi.org/10.1063/5.0005082}
  {\bibfield  {journal} {\bibinfo  {journal} {J. Chem. Phys.}\ }\textbf
  {\bibinfo {volume} {152}},\ \bibinfo {pages} {154105} (\bibinfo {year}
  {2020})}\BibitemShut {NoStop}%
\bibitem [{\citenamefont {Giustino}\ \emph {et~al.}(2007)\citenamefont
  {Giustino}, \citenamefont {Cohen},\ and\ \citenamefont
  {Louie}}]{PhysRevB.76.165108}%
  \BibitemOpen
  \bibfield  {author} {\bibinfo {author} {\bibfnamefont {F.}~\bibnamefont
  {Giustino}}, \bibinfo {author} {\bibfnamefont {M.~L.}\ \bibnamefont
  {Cohen}},\ and\ \bibinfo {author} {\bibfnamefont {S.~G.}\ \bibnamefont
  {Louie}},\ }\bibfield  {title} {\bibinfo {title} {{Electron-phonon
  interaction using Wannier functions}},\ }\href
  {https://doi.org/10.1103/PhysRevB.76.165108} {\bibfield  {journal} {\bibinfo
  {journal} {Phys. Rev. B}\ }\textbf {\bibinfo {volume} {76}},\ \bibinfo
  {pages} {165108} (\bibinfo {year} {2007})}\BibitemShut {NoStop}%
\bibitem [{\citenamefont {Margine}\ and\ \citenamefont
  {Giustino}(2013)}]{PhysRevB.87.024505}%
  \BibitemOpen
  \bibfield  {author} {\bibinfo {author} {\bibfnamefont {E.~R.}\ \bibnamefont
  {Margine}}\ and\ \bibinfo {author} {\bibfnamefont {F.}~\bibnamefont
  {Giustino}},\ }\bibfield  {title} {\bibinfo {title} {{Anisotropic
  Migdal-Eliashberg theory using Wannier functions}},\ }\href
  {https://doi.org/10.1103/PhysRevB.87.024505} {\bibfield  {journal} {\bibinfo
  {journal} {Phys. Rev. B}\ }\textbf {\bibinfo {volume} {87}},\ \bibinfo
  {pages} {024505} (\bibinfo {year} {2013})}\BibitemShut {NoStop}%
\bibitem [{\citenamefont {Ponc{\'e}}\ \emph {et~al.}(2016)\citenamefont
  {Ponc{\'e}}, \citenamefont {Margine}, \citenamefont {Verdi},\ and\
  \citenamefont {Giustino}}]{ponce2016epw}%
  \BibitemOpen
  \bibfield  {author} {\bibinfo {author} {\bibfnamefont {S.}~\bibnamefont
  {Ponc{\'e}}}, \bibinfo {author} {\bibfnamefont {E.~R.}\ \bibnamefont
  {Margine}}, \bibinfo {author} {\bibfnamefont {C.}~\bibnamefont {Verdi}},\
  and\ \bibinfo {author} {\bibfnamefont {F.}~\bibnamefont {Giustino}},\
  }\bibfield  {title} {\bibinfo {title} {{EPW: Electron--phonon coupling,
  transport and superconducting properties using maximally localized Wannier
  functions}},\ }\href
  {https://doi.org/https://doi.org/10.1016/j.cpc.2016.07.028} {\bibfield
  {journal} {\bibinfo  {journal} {Comput. Phys. Commun.}\ }\textbf {\bibinfo
  {volume} {209}},\ \bibinfo {pages} {116} (\bibinfo {year}
  {2016})}\BibitemShut {NoStop}%
\bibitem [{sup()}]{supp}%
  \BibitemOpen
  \href@noop {} {\bibinfo {title} {{See supplementary material:}}},\ \bibinfo
  {note} {the supplementary computational details and theory; phonon specturm
  of NbN bulk, NbN$_2$, Nb$_2$N$_3$, Nb$_3$N$_4$, and Nb$_4$N$_5$ monolayer;
  total electron density of states at the Fermi level N(E$_F$), logarithmic
  frequency average $\omega_\mathrm{log}$ and superconducting critical
  temperature calculated by McMillan-Allen-Dynes equation $T_c$ of NbN bulk,
  NbN$_2$, Nb$_2$N$_3$, Nb$_3$N$_4$, and Nb$_4$N$_5$ monolayer; AIMD
  simulations of the Nb$_2$N$_3$ monolayer; eliashberg spectral function
  $\alpha^2F(\omega)$ and electron phonon coupling $\lambda(\omega)$ by the
  Wannier interpolation; comparison of band structure obtained by density
  functional theory (DFT) and interpolation with maximally localized Wannier
  functions.}\BibitemShut {Stop}%
\bibitem [{\citenamefont {Goldschmid}(2013)}]{goldschmid2013interstitial}%
  \BibitemOpen
  \bibfield  {author} {\bibinfo {author} {\bibfnamefont {H.~J.}\ \bibnamefont
  {Goldschmid}},\ }\href {https://doi.org/10.1007/978-1-4899-5880-8} {\emph
  {\bibinfo {title} {Interstitial alloys}}}\ (\bibinfo  {publisher}
  {Springer},\ \bibinfo {year} {2013})\BibitemShut {NoStop}%
\bibitem [{\citenamefont {Babu}\ and\ \citenamefont
  {Guo}(2019)}]{PhysRevB.99.104508}%
  \BibitemOpen
  \bibfield  {author} {\bibinfo {author} {\bibfnamefont {K.~R.}\ \bibnamefont
  {Babu}}\ and\ \bibinfo {author} {\bibfnamefont {G.-Y.}\ \bibnamefont {Guo}},\
  }\bibfield  {title} {\bibinfo {title} {Electron-phonon coupling,
  superconductivity, and nontrivial band topology in nbn polytypes},\ }\href
  {https://doi.org/10.1103/PhysRevB.99.104508} {\bibfield  {journal} {\bibinfo
  {journal} {Phys. Rev. B}\ }\textbf {\bibinfo {volume} {99}},\ \bibinfo
  {pages} {104508} (\bibinfo {year} {2019})}\BibitemShut {NoStop}%
\bibitem [{\citenamefont {Naguib}\ \emph {et~al.}(2011)\citenamefont {Naguib},
  \citenamefont {Kurtoglu}, \citenamefont {Presser}, \citenamefont {Lu},
  \citenamefont {Niu}, \citenamefont {Heon}, \citenamefont {Hultman},
  \citenamefont {Gogotsi},\ and\ \citenamefont {Barsoum}}]{adma.201102306}%
  \BibitemOpen
  \bibfield  {author} {\bibinfo {author} {\bibfnamefont {M.}~\bibnamefont
  {Naguib}}, \bibinfo {author} {\bibfnamefont {M.}~\bibnamefont {Kurtoglu}},
  \bibinfo {author} {\bibfnamefont {V.}~\bibnamefont {Presser}}, \bibinfo
  {author} {\bibfnamefont {J.}~\bibnamefont {Lu}}, \bibinfo {author}
  {\bibfnamefont {J.}~\bibnamefont {Niu}}, \bibinfo {author} {\bibfnamefont
  {M.}~\bibnamefont {Heon}}, \bibinfo {author} {\bibfnamefont {L.}~\bibnamefont
  {Hultman}}, \bibinfo {author} {\bibfnamefont {Y.}~\bibnamefont {Gogotsi}},\
  and\ \bibinfo {author} {\bibfnamefont {M.~W.}\ \bibnamefont {Barsoum}},\
  }\bibfield  {title} {\bibinfo {title} {{Two-Dimensional Nanocrystals Produced
  by Exfoliation of Ti$_3$AlC$_2$}},\ }\href
  {https://doi.org/https://doi.org/10.1002/adma.201102306} {\bibfield
  {journal} {\bibinfo  {journal} {Adv. Mater.}\ }\textbf {\bibinfo {volume}
  {23}},\ \bibinfo {pages} {4248} (\bibinfo {year} {2011})}\BibitemShut
  {NoStop}%
\bibitem [{\citenamefont {Wang}\ \emph {et~al.}(2018)\citenamefont {Wang},
  \citenamefont {Yu}, \citenamefont {Zhou}, \citenamefont {Li}, \citenamefont
  {Wong}, \citenamefont {Luo}, \citenamefont {Gan},\ and\ \citenamefont
  {Zhai}}]{https://doi.org/10.1002/adfm.201802473}%
  \BibitemOpen
  \bibfield  {author} {\bibinfo {author} {\bibfnamefont {R.}~\bibnamefont
  {Wang}}, \bibinfo {author} {\bibfnamefont {Y.}~\bibnamefont {Yu}}, \bibinfo
  {author} {\bibfnamefont {S.}~\bibnamefont {Zhou}}, \bibinfo {author}
  {\bibfnamefont {H.}~\bibnamefont {Li}}, \bibinfo {author} {\bibfnamefont
  {H.}~\bibnamefont {Wong}}, \bibinfo {author} {\bibfnamefont {Z.}~\bibnamefont
  {Luo}}, \bibinfo {author} {\bibfnamefont {L.}~\bibnamefont {Gan}},\ and\
  \bibinfo {author} {\bibfnamefont {T.}~\bibnamefont {Zhai}},\ }\bibfield
  {title} {\bibinfo {title} {{Strategies on Phase Control in Transition Metal
  Dichalcogenides}},\ }\href {https://doi.org/10.1002/adfm.201802473}
  {\bibfield  {journal} {\bibinfo  {journal} {Adv. Funct. Mater.}\ }\textbf
  {\bibinfo {volume} {28}},\ \bibinfo {pages} {1802473} (\bibinfo {year}
  {2018})}\BibitemShut {NoStop}%
\bibitem [{\citenamefont {Jiang}\ \emph
  {et~al.}(2023{\natexlab{b}})\citenamefont {Jiang}, \citenamefont {Zheng},
  \citenamefont {Kang}, \citenamefont {Tao}, \citenamefont {Huang},
  \citenamefont {Dong},\ and\ \citenamefont {Li}}]{D2TC04799C}%
  \BibitemOpen
  \bibfield  {author} {\bibinfo {author} {\bibfnamefont {P.}~\bibnamefont
  {Jiang}}, \bibinfo {author} {\bibfnamefont {X.}~\bibnamefont {Zheng}},
  \bibinfo {author} {\bibfnamefont {L.}~\bibnamefont {Kang}}, \bibinfo {author}
  {\bibfnamefont {X.}~\bibnamefont {Tao}}, \bibinfo {author} {\bibfnamefont
  {H.-M.}\ \bibnamefont {Huang}}, \bibinfo {author} {\bibfnamefont
  {X.}~\bibnamefont {Dong}},\ and\ \bibinfo {author} {\bibfnamefont {Y.-L.}\
  \bibnamefont {Li}},\ }\bibfield  {title} {\bibinfo {title}
  {{Mn$_2$P$_2$S$_3$Se$_3$: a two-dimensional Janus room-temperature
  antiferromagnetic semiconductor with a large out-of-plane
  piezoelectricity}},\ }\href {https://doi.org/10.1039/D2TC04799C} {\bibfield
  {journal} {\bibinfo  {journal} {J. Mater. Chem. C}\ }\textbf {\bibinfo
  {volume} {11}},\ \bibinfo {pages} {2703} (\bibinfo {year}
  {2023}{\natexlab{b}})}\BibitemShut {NoStop}%
\bibitem [{\citenamefont {Jiang}\ \emph {et~al.}(2020)\citenamefont {Jiang},
  \citenamefont {Kang}, \citenamefont {Zheng}, \citenamefont {Zeng},\ and\
  \citenamefont {Sanvito}}]{PhysRevB.102.195408}%
  \BibitemOpen
  \bibfield  {author} {\bibinfo {author} {\bibfnamefont {P.}~\bibnamefont
  {Jiang}}, \bibinfo {author} {\bibfnamefont {L.}~\bibnamefont {Kang}},
  \bibinfo {author} {\bibfnamefont {X.}~\bibnamefont {Zheng}}, \bibinfo
  {author} {\bibfnamefont {Z.}~\bibnamefont {Zeng}},\ and\ \bibinfo {author}
  {\bibfnamefont {S.}~\bibnamefont {Sanvito}},\ }\bibfield  {title} {\bibinfo
  {title} {{Computational prediction of a two-dimensional semiconductor
  ${\mathrm{SnO}}_{2}$ with negative Poisson's ratio and tunable magnetism by
  doping}},\ }\href {https://doi.org/10.1103/PhysRevB.102.195408} {\bibfield
  {journal} {\bibinfo  {journal} {Phys. Rev. B}\ }\textbf {\bibinfo {volume}
  {102}},\ \bibinfo {pages} {195408} (\bibinfo {year} {2020})}\BibitemShut
  {NoStop}%
\bibitem [{\citenamefont {Haastrup}\ \emph {et~al.}(2018)\citenamefont
  {Haastrup}, \citenamefont {Strange}, \citenamefont {Pandey}, \citenamefont
  {Deilmann}, \citenamefont {Schmidt}, \citenamefont {Hinsche}, \citenamefont
  {Gjerding}, \citenamefont {Torelli}, \citenamefont {Larsen}, \citenamefont
  {Riis-Jensen}, \citenamefont {Gath}, \citenamefont {Jacobsen}, \citenamefont
  {Mortensen}, \citenamefont {Olsen},\ and\ \citenamefont
  {Thygesen}}]{haastrup2018computational}%
  \BibitemOpen
  \bibfield  {author} {\bibinfo {author} {\bibfnamefont {S.}~\bibnamefont
  {Haastrup}}, \bibinfo {author} {\bibfnamefont {M.}~\bibnamefont {Strange}},
  \bibinfo {author} {\bibfnamefont {M.}~\bibnamefont {Pandey}}, \bibinfo
  {author} {\bibfnamefont {T.}~\bibnamefont {Deilmann}}, \bibinfo {author}
  {\bibfnamefont {P.~S.}\ \bibnamefont {Schmidt}}, \bibinfo {author}
  {\bibfnamefont {N.~F.}\ \bibnamefont {Hinsche}}, \bibinfo {author}
  {\bibfnamefont {M.~N.}\ \bibnamefont {Gjerding}}, \bibinfo {author}
  {\bibfnamefont {D.}~\bibnamefont {Torelli}}, \bibinfo {author} {\bibfnamefont
  {P.~M.}\ \bibnamefont {Larsen}}, \bibinfo {author} {\bibfnamefont {A.~C.}\
  \bibnamefont {Riis-Jensen}}, \bibinfo {author} {\bibfnamefont
  {J.}~\bibnamefont {Gath}}, \bibinfo {author} {\bibfnamefont {K.~W.}\
  \bibnamefont {Jacobsen}}, \bibinfo {author} {\bibfnamefont {J.~J.}\
  \bibnamefont {Mortensen}}, \bibinfo {author} {\bibfnamefont {T.}~\bibnamefont
  {Olsen}},\ and\ \bibinfo {author} {\bibfnamefont {K.~S.}\ \bibnamefont
  {Thygesen}},\ }\bibfield  {title} {\bibinfo {title} {{The Computational 2D
  Materials Database: high-throughput modeling and discovery of atomically thin
  crystals}},\ }\href {https://doi.org/10.1088/2053-1583/aacfc1} {\bibfield
  {journal} {\bibinfo  {journal} {2D Mater.}\ }\textbf {\bibinfo {volume}
  {5}},\ \bibinfo {pages} {042002} (\bibinfo {year} {2018})}\BibitemShut
  {NoStop}%
\bibitem [{\citenamefont {Maintz}\ \emph {et~al.}(2016)\citenamefont {Maintz},
  \citenamefont {Deringer}, \citenamefont {Tchougréeff},\ and\ \citenamefont
  {Dronskowski}}]{maintz2016lobster}%
  \BibitemOpen
  \bibfield  {author} {\bibinfo {author} {\bibfnamefont {S.}~\bibnamefont
  {Maintz}}, \bibinfo {author} {\bibfnamefont {V.~L.}\ \bibnamefont
  {Deringer}}, \bibinfo {author} {\bibfnamefont {A.~L.}\ \bibnamefont
  {Tchougréeff}},\ and\ \bibinfo {author} {\bibfnamefont {R.}~\bibnamefont
  {Dronskowski}},\ }\bibfield  {title} {\bibinfo {title} {{LOBSTER: A tool to
  extract chemical bonding from plane-wave based DFT}},\ }\href
  {https://doi.org/https://doi.org/10.1002/jcc.24300} {\bibfield  {journal}
  {\bibinfo  {journal} {J. Comput. Chem.}\ }\textbf {\bibinfo {volume} {37}},\
  \bibinfo {pages} {1030} (\bibinfo {year} {2016})}\BibitemShut {NoStop}%
\bibitem [{\citenamefont {Shan}\ \emph {et~al.}(2025)\citenamefont {Shan},
  \citenamefont {Ma}, \citenamefont {Yang}, \citenamefont {Li}, \citenamefont
  {Liu}, \citenamefont {Hou}, \citenamefont {Jiang}, \citenamefont {Zhang},
  \citenamefont {Shi}, \citenamefont {Yang}, \citenamefont {Lin}, \citenamefont
  {Wang}, \citenamefont {Sun}, \citenamefont {Guo}, \citenamefont {Ding},
  \citenamefont {Gou}, \citenamefont {Zhao},\ and\ \citenamefont
  {Cheng}}]{doi:10.1021/jacs.4c15137}%
  \BibitemOpen
  \bibfield  {author} {\bibinfo {author} {\bibfnamefont {P.}~\bibnamefont
  {Shan}}, \bibinfo {author} {\bibfnamefont {L.}~\bibnamefont {Ma}}, \bibinfo
  {author} {\bibfnamefont {X.}~\bibnamefont {Yang}}, \bibinfo {author}
  {\bibfnamefont {M.}~\bibnamefont {Li}}, \bibinfo {author} {\bibfnamefont
  {Z.}~\bibnamefont {Liu}}, \bibinfo {author} {\bibfnamefont {J.}~\bibnamefont
  {Hou}}, \bibinfo {author} {\bibfnamefont {S.}~\bibnamefont {Jiang}}, \bibinfo
  {author} {\bibfnamefont {L.}~\bibnamefont {Zhang}}, \bibinfo {author}
  {\bibfnamefont {L.}~\bibnamefont {Shi}}, \bibinfo {author} {\bibfnamefont
  {P.}~\bibnamefont {Yang}}, \bibinfo {author} {\bibfnamefont {C.}~\bibnamefont
  {Lin}}, \bibinfo {author} {\bibfnamefont {B.}~\bibnamefont {Wang}}, \bibinfo
  {author} {\bibfnamefont {J.}~\bibnamefont {Sun}}, \bibinfo {author}
  {\bibfnamefont {H.}~\bibnamefont {Guo}}, \bibinfo {author} {\bibfnamefont
  {Y.}~\bibnamefont {Ding}}, \bibinfo {author} {\bibfnamefont {H.}~\bibnamefont
  {Gou}}, \bibinfo {author} {\bibfnamefont {Z.}~\bibnamefont {Zhao}},\ and\
  \bibinfo {author} {\bibfnamefont {J.}~\bibnamefont {Cheng}},\ }\bibfield
  {title} {\bibinfo {title} {{Molecular Hydride Superconductor BiH$_4$ with
  $T_c$ up to 91 K at 170 GPa}},\ }\href {https://doi.org/10.1021/jacs.4c15137}
  {\bibfield  {journal} {\bibinfo  {journal} {J. Am. Chem. Soc.}\ }\textbf
  {\bibinfo {volume} {147}},\ \bibinfo {pages} {4375} (\bibinfo {year}
  {2025})}\BibitemShut {NoStop}%
\bibitem [{\citenamefont {Deng}\ \emph {et~al.}(2024)\citenamefont {Deng},
  \citenamefont {Wang}, \citenamefont {Huang}, \citenamefont {Xu},
  \citenamefont {Zhao}, \citenamefont {Du}, \citenamefont {Song},\ and\
  \citenamefont {Cui}}]{PhysRevB.109.184516}%
  \BibitemOpen
  \bibfield  {author} {\bibinfo {author} {\bibfnamefont {C.}~\bibnamefont
  {Deng}}, \bibinfo {author} {\bibfnamefont {M.}~\bibnamefont {Wang}}, \bibinfo
  {author} {\bibfnamefont {H.}~\bibnamefont {Huang}}, \bibinfo {author}
  {\bibfnamefont {M.}~\bibnamefont {Xu}}, \bibinfo {author} {\bibfnamefont
  {W.}~\bibnamefont {Zhao}}, \bibinfo {author} {\bibfnamefont {M.}~\bibnamefont
  {Du}}, \bibinfo {author} {\bibfnamefont {H.}~\bibnamefont {Song}},\ and\
  \bibinfo {author} {\bibfnamefont {T.}~\bibnamefont {Cui}},\ }\bibfield
  {title} {\bibinfo {title} {{High-${T}_{c}$ superconductors in the ternary
  Sr-Hf/Zr-H system at high pressure}},\ }\href
  {https://doi.org/10.1103/PhysRevB.109.184516} {\bibfield  {journal} {\bibinfo
   {journal} {Phys. Rev. B}\ }\textbf {\bibinfo {volume} {109}},\ \bibinfo
  {pages} {184516} (\bibinfo {year} {2024})}\BibitemShut {NoStop}%
\bibitem [{\citenamefont {Modak}\ \emph {et~al.}(2021)\citenamefont {Modak},
  \citenamefont {Verma},\ and\ \citenamefont {Mishra}}]{PhysRevB.104.054504}%
  \BibitemOpen
  \bibfield  {author} {\bibinfo {author} {\bibfnamefont {P.}~\bibnamefont
  {Modak}}, \bibinfo {author} {\bibfnamefont {A.~K.}\ \bibnamefont {Verma}},\
  and\ \bibinfo {author} {\bibfnamefont {A.~K.}\ \bibnamefont {Mishra}},\
  }\bibfield  {title} {\bibinfo {title} {{Prediction of superconductivity at 70
  K in a pristine monolayer of LiBC}},\ }\href
  {https://doi.org/10.1103/PhysRevB.104.054504} {\bibfield  {journal} {\bibinfo
   {journal} {Phys. Rev. B}\ }\textbf {\bibinfo {volume} {104}},\ \bibinfo
  {pages} {054504} (\bibinfo {year} {2021})}\BibitemShut {NoStop}%
\bibitem [{\citenamefont {McMillan}(1968)}]{PhysRev.167.331}%
  \BibitemOpen
  \bibfield  {author} {\bibinfo {author} {\bibfnamefont {W.~L.}\ \bibnamefont
  {McMillan}},\ }\bibfield  {title} {\bibinfo {title} {{Transition Temperature
  of Strong-Coupled Superconductors}},\ }\href
  {https://doi.org/10.1103/PhysRev.167.331} {\bibfield  {journal} {\bibinfo
  {journal} {Phys. Rev.}\ }\textbf {\bibinfo {volume} {167}},\ \bibinfo {pages}
  {331} (\bibinfo {year} {1968})}\BibitemShut {NoStop}%
\bibitem [{\citenamefont {Dynes}(1972)}]{DYNES1972615}%
  \BibitemOpen
  \bibfield  {author} {\bibinfo {author} {\bibfnamefont {R.}~\bibnamefont
  {Dynes}},\ }\bibfield  {title} {\bibinfo {title} {{McMillan's equation and
  the $T_c$ of superconductors}},\ }\href
  {https://doi.org/https://doi.org/10.1016/0038-1098(72)90603-5} {\bibfield
  {journal} {\bibinfo  {journal} {Solid State Commun.}\ }\textbf {\bibinfo
  {volume} {10}},\ \bibinfo {pages} {615} (\bibinfo {year} {1972})}\BibitemShut
  {NoStop}%
\bibitem [{\citenamefont {Allen}\ and\ \citenamefont
  {Dynes}(1975)}]{PhysRevB.12.905}%
  \BibitemOpen
  \bibfield  {author} {\bibinfo {author} {\bibfnamefont {P.~B.}\ \bibnamefont
  {Allen}}\ and\ \bibinfo {author} {\bibfnamefont {R.~C.}\ \bibnamefont
  {Dynes}},\ }\bibfield  {title} {\bibinfo {title} {{Transition temperature of
  strong-coupled superconductors reanalyzed}},\ }\href
  {https://doi.org/10.1103/PhysRevB.12.905} {\bibfield  {journal} {\bibinfo
  {journal} {Phys. Rev. B}\ }\textbf {\bibinfo {volume} {12}},\ \bibinfo
  {pages} {905} (\bibinfo {year} {1975})}\BibitemShut {NoStop}%
\bibitem [{\citenamefont {Bekaert}\ \emph {et~al.}(2020)\citenamefont
  {Bekaert}, \citenamefont {Sevik},\ and\ \citenamefont
  {Milošević}}]{D0NR03875J}%
  \BibitemOpen
  \bibfield  {author} {\bibinfo {author} {\bibfnamefont {J.}~\bibnamefont
  {Bekaert}}, \bibinfo {author} {\bibfnamefont {C.}~\bibnamefont {Sevik}},\
  and\ \bibinfo {author} {\bibfnamefont {M.~V.}\ \bibnamefont {Milošević}},\
  }\bibfield  {title} {\bibinfo {title} {First-principles exploration of
  superconductivity in mxenes},\ }\href {https://doi.org/10.1039/D0NR03875J}
  {\bibfield  {journal} {\bibinfo  {journal} {Nanoscale}\ }\textbf {\bibinfo
  {volume} {12}},\ \bibinfo {pages} {17354} (\bibinfo {year}
  {2020})}\BibitemShut {NoStop}%
\bibitem [{\citenamefont {Zha}\ \emph {et~al.}(2023)\citenamefont {Zha},
  \citenamefont {Jiang}, \citenamefont {Huang},\ and\ \citenamefont
  {Li}}]{PhysRevMaterials.7.114805}%
  \BibitemOpen
  \bibfield  {author} {\bibinfo {author} {\bibfnamefont {D.-B.}\ \bibnamefont
  {Zha}}, \bibinfo {author} {\bibfnamefont {P.}~\bibnamefont {Jiang}}, \bibinfo
  {author} {\bibfnamefont {H.-M.}\ \bibnamefont {Huang}},\ and\ \bibinfo
  {author} {\bibfnamefont {Y.-L.}\ \bibnamefont {Li}},\ }\bibfield  {title}
  {\bibinfo {title} {{Refined phase diagram and kagome-lattice
  superconductivity in Mg-Si system}},\ }\href
  {https://doi.org/10.1103/PhysRevMaterials.7.114805} {\bibfield  {journal}
  {\bibinfo  {journal} {Phys. Rev. Mater.}\ }\textbf {\bibinfo {volume} {7}},\
  \bibinfo {pages} {114805} (\bibinfo {year} {2023})}\BibitemShut {NoStop}%
\bibitem [{\citenamefont {Choi}\ \emph {et~al.}(2002)\citenamefont {Choi},
  \citenamefont {Roundy}, \citenamefont {Sun}, \citenamefont {Cohen},\ and\
  \citenamefont {Louie}}]{PhysRevB.66.020513}%
  \BibitemOpen
  \bibfield  {author} {\bibinfo {author} {\bibfnamefont {H.~J.}\ \bibnamefont
  {Choi}}, \bibinfo {author} {\bibfnamefont {D.}~\bibnamefont {Roundy}},
  \bibinfo {author} {\bibfnamefont {H.}~\bibnamefont {Sun}}, \bibinfo {author}
  {\bibfnamefont {M.~L.}\ \bibnamefont {Cohen}},\ and\ \bibinfo {author}
  {\bibfnamefont {S.~G.}\ \bibnamefont {Louie}},\ }\bibfield  {title} {\bibinfo
  {title} {{First-principles calculation of the superconducting transition in
  ${\mathrm{MgB}}_{2}$ within the anisotropic Eliashberg formalism}},\ }\href
  {https://doi.org/10.1103/PhysRevB.66.020513} {\bibfield  {journal} {\bibinfo
  {journal} {Phys. Rev. B}\ }\textbf {\bibinfo {volume} {66}},\ \bibinfo
  {pages} {020513} (\bibinfo {year} {2002})}\BibitemShut {NoStop}%
\bibitem [{\citenamefont {Fine}(2005)}]{PhysRevLett.94.157005}%
  \BibitemOpen
  \bibfield  {author} {\bibinfo {author} {\bibfnamefont {B.~V.}\ \bibnamefont
  {Fine}},\ }\bibfield  {title} {\bibinfo {title} {Temperature dependence of
  the superconducting gap in high-${T}_{c}$ cuprates},\ }\href
  {https://doi.org/10.1103/PhysRevLett.94.157005} {\bibfield  {journal}
  {\bibinfo  {journal} {Phys. Rev. Lett.}\ }\textbf {\bibinfo {volume} {94}},\
  \bibinfo {pages} {157005} (\bibinfo {year} {2005})}\BibitemShut {NoStop}%
\bibitem [{\citenamefont {Clem}(1966)}]{CLEM1966268}%
  \BibitemOpen
  \bibfield  {author} {\bibinfo {author} {\bibfnamefont {J.~R.}\ \bibnamefont
  {Clem}},\ }\bibfield  {title} {\bibinfo {title} {Effects of energy gap
  anisotropy in pure superconductors},\ }\href
  {https://doi.org/https://doi.org/10.1016/0003-4916(66)90028-5} {\bibfield
  {journal} {\bibinfo  {journal} {Annals of Physics}\ }\textbf {\bibinfo
  {volume} {40}},\ \bibinfo {pages} {268} (\bibinfo {year} {1966})}\BibitemShut
  {NoStop}%
\end{thebibliography}%

\end{document}